\documentclass[11pt]{article}

\usepackage{amsfonts,amssymb,amsmath,geometry}
\usepackage{graphicx,color}
\usepackage{subfigure}
\usepackage{afterpage}

\begin{document}
\def\singlespacing{\baselineskip=12pt}	\def\doublespacing{\baselineskip=18pt}
\singlespacing

\pagestyle{myheadings}\markright{{\sc Dynamic effective medium theories for periodic  media}  ~~~~~~\today}

\markboth{3D dynamic homogenization}{3D dynamic homogenization}
\title{Analytical formulation of 3D dynamic homogenization for periodic elastic systems }
\author{A.N. Norris$^{a}$, A.L. Shuvalov$^{b}$, A.A. Kutsenko$^{b}$  \\ \\
$^{a}$\ Mechanical and Aerospace Engineering, Rutgers University,\\
Piscataway, NJ 08854-8058, USA \\
$^{b}$ Universit\'{e} de Bordeaux, CNRS, UMR 5469, \\
Laboratoire de M\'{e}canique Physique, Talence 33405, France
}


\maketitle

\begin{abstract}

Homogenization of the equations of motion for a three dimensional periodic elastic system is considered. Expressions are obtained for the fully dynamic  effective material parameters governing the spatially averaged
fields by using the plane wave expansion (PWE) method.  The effective equations are of Willis form \cite{Willis97} with coupling between momentum and stress and tensorial inertia.  The formulation  demonstrates that the Willis equations of elastodynamics are closed under homogenization. The  effective material parameters are obtained for  arbitrary frequency and wavenumber combinations, including but not restricted to  Bloch wave branches for wave propagation in the periodic medium.  Numerical examples for a 1D system illustrate the frequency dependence of the parameters on Bloch wave branches and provide a comparison with an alternative  dynamic effective medium theory \cite{Shuvalov11} which also reduces to Willis form but with different effective moduli.
\end{abstract}

%

\section{Introduction}

\label{sec1}

The effective medium concept has proven to be very useful for modelling
composite materials with periodic microstructure. 
Efficient methods exist for computing the 
effective elastic moduli under static and quasistatic conditions. A more challenging task is to define frequency-dependent dynamic effective constants which are capable of describing phononic band gaps of the wave spectrum \cite{Liu05} at the expense of non-classical features such as negative \cite{Avila08} and tensorial \cite{Milton07} density. Effective material models that can retain a dispersive signature of the initial material microarchitecture
have significant potential application in the design
of sonic metamaterials. The purpose of this paper is to provide an
analytical formulation for the effective material parameters of a periodic
elastic medium at finite frequency.

A general framework for describing dynamic effective medium theories has
been developed by Willis, beginning with variational arguments for non-local
behavior in mechanical \cite{Willis81a,Willis81b} and electromagnetic
systems \cite{Willis84}. The predicted constitutive relations modify both the
elasticity \cite{Willis83} and the inertia \cite{Willis85}, and include
coupling between stress/strain on the one hand and momentum/velocity on the
other, see \cite{Willis97} for a review. The notion of anisotropic inertia
has appeared in several mechanical contexts, such as stratified layers of
fluids \cite{Schoenberg83} and elastic composites with strongly contrasting
constituents \cite{Avila05}. Coupling between particle velocity and stress,
although not usually considered in constitutive theories for continua, is to
be expected for theories relating spatial means of these quantities because
of the underlying inhomogeneity of the displacement and stress fields within
the region being averaged. Interest in the Willis constitutive model has
intensified with the observation \cite{Milton07} that the non-classical
material properties could be realized, in theory at least, by suitably
designed microstructures with internal springs, masses, and gyroscopes.
Derivations of the Willis equations have been demonstrated for periodic
systems in 1D \cite{Willis09,Shuvalov11,Nasser11a} 
and 3D \cite%
{Nasser11b}, and the structure of the equations has been rigorously proved
for both electromagnetic and elastic materials \cite{Willis11}.

Despite the growing awareness that the Willis constitutive model is the
proper dynamic effective medium formulation for periodic systems there is as
yet no simple means of calculating the effective material parameters at
finite frequency. Several different techniques have been proposed to address
this deficiency, none of which is easily implemented for arbitrary 3D
problems. Thus, closed form expressions were obtained in \cite{Willis09} for
1D stratification using Floquet modes, where the latter have to be
determined numerically. A more general numerical technique has recently been
developed and applied to 1D \cite{Nasser11a} and 3D \cite{Nasser11b}
composites. This approach employs a background or reference material, and
solves for the polarization field (difference in stress/momentum between the
actual and that of a comparison body \cite{Willis1980}) to arrive at
expressions for the effective parameters in terms of system matrices. The
numerical procedure, which uses a combination of plane wave expansions and
discretization schemes, is  complicated by the required
selection of a comparison material, the choice of which has no bearing on
the final outcome. A quite different technique \cite{Shuvalov11} suited to
1D-periodically stratified elastic solids uses the monodromy matrix (i.e.,
propagator  for a single period) to explicitly define effective material
parameters from its logarithm. All of these methods derive
the density, stiffness and the coupling parameter, $S$, as functions of
frequency and wavenumber.  
This means, perhaps surprisingly, that
equations of Willis form but with different parameters can be obtained for
the same medium, depending on the homogenization scheme employed, an aspect
discussed in \S \ref{sec6}.

Our purpose here is to develop, in so far as possible, analytical
expressions for the dynamic effective material parameters of a periodic
elastic material. The homogenization scheme seeks the constitutive relations
and the equations of dynamic equilibrium governing the spatial average of
the \textit{periodic} part of the Bloch wave solution. 
The initial heterogeneous system is here assumed to be defined by pointwise constitutive
equations of the form proposed by Willis, which includes 'classical'
elastodynamics as the case of isotropic density and zero coupling between
stress and momentum. We show that the homogenized equations are also of
Willis form with dynamic effective material parameters that are non-local in
space and time, thus demonstrating that the Willis constitutive theory is
self-consistent and closed. The homogenized effective parameters are
conveniently expressed in terms of plane wave expansions (PWE) of the
original material parameters, and accordingly we call this the PWE method. 
{This approach allows us to define
  effective constants for any frequency-wavenumber combination, including, but not restricted to, values of 
	$\{ \omega,\mathbf {k}\}$  on the Bloch wave branches. }

We begin in \S \ref{sec2} with an overview of the Willis constitutive
equations followed by a summary of the PWE homogenization results. 
Section \S \ref{sec3} is devoted to the derivation of the PWE effective
material parameters for a scalar wave model that is simpler than the fully
elastic problem, but which exhibits the main features of the homogenization
scheme. The elastodynamic effective parameters are obtained in \S \ref{sec4}
and some of their properties are examined in \S \ref{sec5}. Examples of
the PWE effective moduli are provided in \S \ref{sec6} based on numerical
calculations for a 1D system, and comparisons are made with the parameters
predicted by the monodromy matrix (MM) homogenization scheme \cite{Shuvalov11}. 
Conclusions and final words are given in \S \ref{sec7}.

\section{Problem setup  and summary of the solution}

\label{sec2}

\subsection{Willis elastodynamic equations for a heterogeneous medium}

\label{2.1}

The field variables are the vectors of displacement, momentum and body
force, ${\mathbf{u}}$, ${\mathbf{p}}$, and ${\mathbf{f}}$; the symmetric
second order tensors of stress, strain, and strain source, ${\pmb\sigma }$, $%
{\pmb\varepsilon}$, ${\pmb\gamma }$, with ${\pmb\varepsilon}=\tfrac{1}{2}[%
\mathbf{\nabla }{\mathbf{u}}+(\mathbf{\nabla }{\mathbf{u}})^{T}]$. They are
related by the system of elastodynamic equations which we  take in the
Willis form (in order to demonstrate in the end that this model is closed
under homogenization for periodic materials). Accordingly, the material
parameters are the second order tensor of mass density ${\pmb\rho }$, the
fourth order elastic stiffness tensor ${\pmb C,}$ and a third order coupling
tensor ${\pmb S}$. The components of the above material tensors in an
orthogonal basis of the coordinate space satisfy the symmetries 
\begin{equation}
\rho _{ij}=\rho _{ji}^{\ast },\quad C_{ijkl}=C_{klij}^{\ast },\quad
C_{ijkl}=C_{jikl},\quad S_{ijk}=S_{jik}.  \label{-434}
\end{equation}%
The two former identities imply that the density and stiffness tensors are
Hermitian: ${\pmb\rho } ={\pmb\rho }^{+}$ and $\mathbf{C=C}^{+}$ ($^{\ast }$ and 
$^{+}$ hereafter denote complex and Hermitian conjugation). Assuming both $%
{\pmb\rho }$ and $\mathbf{C}$ positive definite ensures positive energy (%
\ref{=65}), see below. Let ${\mathbf{S}}$ be generally complex. Field
variables are functions of position $\mathbf{x}$ and time $t$, while the
material tensors in the initial heterogeneous medium are considered as
functions of $\mathbf{x}$ only (the homogenization  leads to the PWE
effective parameters which, after inverse Fourier transform, are functions
of both $\mathbf{x}$ and $t$).

The equation of equilibrium and the generalized constitutive equations in
Willis form \cite{Milton07,Willis11} are, respectively, 
\begin{equation}
\dot{\mathbf{p}}=\text{div}\mathbf{\sigma }+\mathbf{f},\qquad 
\begin{pmatrix}
{\pmb\sigma } \\ 
\\ 
{\mathbf{p}}%
\end{pmatrix}%
=%
\begin{pmatrix}
{\mathbf{C}} & {\mathbf{S}} \\ 
&  \\ 
-{\mathbf{S}}^{+} & {\pmb\rho }%
\end{pmatrix}%
\begin{pmatrix}
{\pmb\varepsilon}-{\pmb\gamma } \\ 
\\ 
\dot{\mathbf{u}}%
\end{pmatrix}%
,  \label{=64}
\end{equation}%
where the dot implies time derivative. Note that the terms of (\ref{=64})
with $\mathbf{S}$ taken in component form imply ${\sigma }_{ij}\ni S_{\left(
ij\right) k}\dot{u}_{k}$ and $p_{i}\ni -{S}_{i\left( jk\right) }^{+}\left( {%
\varepsilon -\gamma }\right) _{jk}=-{S}_{\left( jk\right) i}^{\ast }\left( {%
\varepsilon -\gamma }\right) _{jk}$, where the permutable indices are
enclosed in parentheses  (see  \S \ref{4.1} for further details). 
 By
(\ref{-434}) and (\ref{=64}) with $\mathbf{\gamma =0,}$ the time-averaged
energy density is 
\begin{equation}
E=\frac{1}{2}\text{Re}({\pmb\sigma }^{+}{\pmb\varepsilon}+{\dot{\mathbf{u}}}%
^{+}{\mathbf{p}})=\frac{1}{2}\left( {\pmb\varepsilon}^{+}
\mathbf{C}{\pmb\varepsilon} +\dot{\mathbf{u}}^{+} {\pmb\rho } \dot{\mathbf{u}}
\right) .
\label{=65}
\end{equation}%
It is obvious that eqs.\ \eqref{=64} include 'classical' linearly
anisotropic elasticity with $\rho _{ij}=\rho \delta _{ij}$, $%
C_{ijkl}=C_{klij}=C_{jikl}$ and $S_{ijl}=0$.

Equations \eqref{=64} may be solved to find ${\mathbf{u}}$, ${\mathbf{p}}$,
and ${\pmb\sigma }$ for given forcing functions ${\mathbf{f}}$ and ${\pmb%
\gamma }$. Accounting for the effect of two driving-force terms is an
important ingredient of homogenization concepts \cite{Willis11}. In
particular, dependence of the solutions on the forcing terms will turn out
to be crucial in finding the PWE homogenized equations.

\subsection{The periodic medium}

We consider the material with $\mathbf{T}$-periodic parameters: 
\begin{equation}
h\big(\mathbf{x+}%
\begin{matrix}
\sum\nolimits_{j=1}^{d}%
\end{matrix}%
n_{j}\mathbf{a}_{j}\big)=h(\mathbf{x})\quad \text{for }\ h={\mathbf{C}},{\pmb%
\rho },{\mathbf{S}},  \label{19}
\end{equation}%
where $\mathbf{x}\in \mathbb{R}^{d}$ ($d=$1, 2 or 3), $n_{j}\in \mathbb{Z}$,
and the linear independent translation vectors $\mathbf{a}_{j}\in \mathbb{R}%
^{d}$ define the irreducible unit cell $\mathbf{T}=\sum%
\nolimits_{j=1}^{d}t_{j}\mathbf{a}_{j}$ ($t_{j}\in \left[ 0,1\right] $) of
the periodic lattice. Let $\left\{ \mathbf{e}_{j}\right\} $ with $j=1,\ldots
,d$ be an orthonormal base in $\mathbb{R}^{d}$. Denote 
\begin{equation}
\mathbf{a}_{j}=\mathbf{Ae}_{j},\ \mathbf{b}_{j}=\left( \mathbf{A}%
^{-1}\right) ^{\mathrm{T}}\mathbf{e}_{j},\quad \Gamma =\{\mathbf{g}:\ 
\mathbf{g}=%
\begin{matrix}
\sum\nolimits_{j=1}^{d}%
\end{matrix}%
2\pi n_{j}\mathbf{b}_{j},\ n_{j}\in \mathbb{Z}\},  \label{101}
\end{equation}%
where $\mathbf{a}_{j}\cdot \mathbf{b}_{k}=\delta _{jk}$ ($\cdot $ is the
scalar product in $\mathbb{R}^{d}$), \thinspace\ $^{\mathrm{T}}$ means
transpose, and $\mathbf{g}=\sum\nolimits_{j=1}^{d}g_{j}\mathbf{e}_{j}$ is an
element of the set $\Gamma $ of reciprocal lattice vectors whose 
components $\left\{ g_{j}\right\} $ in the base $\left\{ \mathbf{e}_{j}\right\} $ take all values $2\pi \left( \mathbf{A}%
^{-1}\right) ^{\mathrm{T}}\mathbb{Z}^{d}$. 
The Fourier transform maps a $\mathbf{T}$-periodic function $h(\mathbf{x})$
to a vector $\hat{\mathbf{h}}$ in the infinite-dimensional space $V_{\mathbf{%
g}}$ associated with $\mathbf{g}\in \Gamma ,$ where the components $\hat{h}(%
\mathbf{g})$ of $\hat{\mathbf{h}}$ are Fourier coefficients of $h(\mathbf{x}%
):$ 
\begin{equation}
\hat{h}(\mathbf{g})=|\mathbf{T}|^{-1}\int_{0}^{\mathbf{T}}h(\mathbf{x})e^{-i%
\mathbf{g}\cdot \mathbf{x}}\text{d}\mathbf{x}\quad \Leftrightarrow \quad h(%
\mathbf{x})=%
\begin{matrix}
\sum\nolimits_{\mathbf{g}\in \Gamma }%
\end{matrix}%
\hat{h}(\mathbf{g})e^{i\mathbf{g}\cdot \mathbf{x}}  \label{3-5}
\end{equation}%
(all quantities in the Fourier domain are indicated by a hat hereafter). In
particular, the average over the single cell is defined by%
\begin{equation}
\langle h\rangle \equiv |\mathbf{T}|^{-1}\int_{0}^{\mathbf{T}}h(\mathbf{x})%
\text{d}\mathbf{x}\ \ \big(=\hat{h}(0)=\hat{\mathbf{e}}^{+}\hat{\mathbf{h}}%
\mathrm{\quad where\ \ }\hat{e}\left( \mathbf{g}\right) =\delta _{\mathbf{g}%
\mathbf{0}}\big).  \label{3-4}
\end{equation}%
Practical calculations in the Fourier domain imply truncation of vectors and
matrices in $V_{\mathbf{g}}$ to a finite size, i.e.\ reducing $V_{\mathbf{g}}$
to finite dimension. This is tacitly understood in the following when using
the concepts of matrix determinant and inverse as convenient shortcuts.

The governing equations (\ref{=64}) with periodic coefficients (\ref{19})
will be solved for both the wave field variables and the force terms in
Bloch form 
\begin{equation}
h(\mathbf{x},t)=h(\mathbf{x})e^{i(\mathbf{k}\cdot \mathbf{x}-\omega
t)},\quad h(\mathbf{x})\ \ \mathbf{T}\mathrm{-periodic,\quad for\ }h={%
\mathbf{u}},{\mathbf{p}},{\pmb\sigma },{\pmb\varepsilon},{\mathbf{f}},{\pmb%
\gamma ,}  \label{2}
\end{equation}%
where $h(\mathbf{x})$ implies \textit{the periodic part} of the full
dependence of $h$ on $\mathbf{x}$ (this convention reduces subsequent
notation), and the frequency $\omega $ and wave-vector $\mathbf{k}$ are
assumed real-valued. Note that one may think of ${\mathbf{f}}$ and ${\pmb%
\gamma } $ in the form (\ref{2}) as phased forcing terms which 'drive' the
solution.

\subsection{Summary of PWE homogenization results}

\label{sum1}

For $h$ expressed in Bloch wave form, \eqref{2}, define the effective field
variable 
\begin{equation}
h^{\text{eff}}(\mathbf{x},t)=\langle h\rangle \,e^{i(\mathbf{k}\cdot \mathbf{%
x}-\omega t)}.  \label{3-2}
\end{equation}%
Using the effective sources $\left( \mathbf{f}^{\text{eff}},\mathbf{\gamma }%
^{\text{eff}}\right) =\left( \langle \mathbf{f}\rangle ,\left\langle \mathbf{%
\gamma }\right\rangle \right) \,e^{i(\mathbf{k}\cdot \mathbf{x}-\omega t)}$
implies ignoring the influence from the forcing $\mathbf{f}(\mathbf{x})$, ${%
\pmb\gamma }(\mathbf{x})$ with zero average over a single cell, which is
natural for the homogenization theory aimed at recovering the effective wave
fields of the form (\ref{3-2}). One of the principal results of the paper is
that, given the periodic medium with the Willis form (\ref{=64}) of the
governing equations (which incorporates 'classical' elastodynamics model),
the equations describing relations between the effective wave fields ${%
\mathbf{u}}^{\text{eff}}$, ${\pmb\sigma }^{\text{eff}}$, ${\mathbf{p}}^{%
\text{eff}}$ and ${\pmb\varepsilon}^{\text{eff}}\ =\tfrac{1}{2}[\mathbf{%
\nabla }{\mathbf{u}}^{\text{eff}}\ +(\mathbf{\nabla }{\pmb u}^{\text{eff}%
})^{T}]$ and the effective forcing ${\mathbf{f}}^{\text{eff}}$ and ${\pmb%
\gamma }^{\text{eff}}$ are also of Willis form: 
\begin{equation}
\dot{\mathbf{p}}^{\text{eff}}\ =\text{div}\mathbf{\sigma }^{\text{eff}}\ +%
\mathbf{f}^{\text{eff}},\qquad 
\begin{pmatrix}
\mathbf{\sigma }^{\text{eff}} \\ 
\\ 
\mathbf{p}^{\text{eff}}%
\end{pmatrix}%
=%
\begin{pmatrix}
\mathbf{C}^{\text{eff}} & \mathbf{S}^{\text{eff}} \\ 
&  \\ 
-{\mathbf{S}^{\text{eff}}}^{+} & {\pmb\rho }^{\text{eff}}%
\end{pmatrix}%
\begin{pmatrix}
{\pmb\epsilon }^{\text{eff}}\ -{\pmb\gamma }^{\text{eff}} \\ 
\\ 
\dot{\mathbf{u}}^{\text{eff}}%
\end{pmatrix}%
.  \label{00}
\end{equation}%
The effective parameters ${\mathbf{C}}^{\text{eff}}$, ${\pmb\rho }^{\text{eff%
}}$ and ${\mathbf{S}}^{\text{eff}}$ are non-local in both space and time
(despite the fact that the exact parameters of the original periodic
material are stationary). Correspondingly, they are functions of frequency $%
\omega $ and wavenumber vector $\mathbf{k}$ in the transform domain ($\frac{%
\partial }{\partial x_{j}},\frac{\partial }{\partial t}\rightarrow
ik_{j},-i\omega $), where their explicit expressions are as follows: 
\begin{subequations}
\label{-13}
\begin{align}
C_{ijkl}^{\text{eff}}(\omega ,\mathbf{k})& =\langle C_{ijkl}\rangle -\hat{%
\mathbf{q}}_{ijp}^{+}\hat{\mathbf{G}}_{pq}^{\!\overset{\backslash \!\mathbf{0%
}}{}}\hat{\mathbf{q}}_{klq},  \label{-3a} \\
S_{ijk}^{\text{eff}}(\omega ,\mathbf{k})& =\langle S_{ijk}\rangle -\hat{%
\mathbf{q}}_{ijp}^{+}\hat{\mathbf{G}}_{pq}^{\!\overset{\backslash \!\mathbf{0%
}}{}}\hat{\mathbf{r}}_{{q k}},  \label{-3b} \\
\rho _{ik}^{\text{eff}}(\omega ,\mathbf{k})& =\langle \rho _{ik}\rangle +%
\hat{\mathbf{r}}_{{pi}}^{+}\hat{\mathbf{G}}_{pq}^{\!\overset{%
\backslash \!\mathbf{0}}{}}\hat{\mathbf{r}}_{{q k}},
\label{-3c}
\end{align}%
with the summation on repeated indices hereafter implicitly understood.
(Noting that the indices of tensors ${\mathbf{C}}$ and ${\mathbf{S}}$ are
appropriately split between $\mathbb{C}^{3}\ni \mathbf{u}$ and $\mathbb{R}%
^{d}\ni \mathbf{x}$, we assume for brevity that
$d=3$ and so all
roman indices run from 1 to 3). The tensors $\hat{\mathbf{q}}$, $\hat{%
\mathbf{r}}~$and $\hat{\mathbf{G}}^{\backslash \mathbf{0}}$ are defined on $%
V_{\mathbf{g\neq 0}}\otimes \mathbb{C}^{3}$ where $V_{\mathbf{g\neq 0}}$ is
the Fourier vector space restricted to ${\mathbf{g}\in \Gamma \backslash 
\mathbf{0.}}$ Each 'element' $\hat{\mathbf{q}}_{ijk}$, $\hat{\mathbf{r}}%
_{ik}~$and $\hat{\mathbf{G}}_{pq}^{\!\overset{\backslash \!\mathbf{0}}{}}$
with fixed indices $i...q$ is, respectively, a vector and a matrix in $V_{%
\mathbf{g\neq 0}}.$ Their components are as follows:%
\begin{align}
\hat{q}_{ijk}(\mathbf{g})& =(g_{l}+k_{l})\hat{C}_{ijkl}(\mathbf{g})-\omega 
\hat{S}_{ijk}^{\ast }(\mathbf{g}),\ \ \ \ {\mathbf{g}\in \Gamma \backslash 
\mathbf{0}};  \label{-72} \\
\hat{r}_{ik}(\mathbf{g})& =\omega \hat{\rho}_{ik}(\mathbf{g})+(g_{j}+k_{j})%
\hat{S}_{ijk}(\mathbf{g}),\quad {\mathbf{g}\in \Gamma \backslash \mathbf{0}};
\\
\sum\nolimits_{\mathbf{g}^{\prime }}\hat{G}_{ip}^{\overset{\backslash \!%
\mathbf{0}}{}}\left[ \mathbf{g},\mathbf{g}^{\prime }\right] \hat{Z}_{pk}^{%
\overset{\backslash \!\mathbf{0}}{}}\left[ \mathbf{g}^{\prime },\mathbf{g}%
^{\prime \prime }\right] & =\delta _{ik}\delta _{\mathbf{g}\mathbf{g}%
^{\prime \prime }}\mathrm{\quad with\ }{\mathbf{g},\,}\mathbf{g}^{\prime },\,%
\mathbf{g}^{\prime \prime }{\in \Gamma \backslash \mathbf{0}},\ \ \text{and}
\label{-72f} \\
\hat{Z}_{ik}^{\overset{\backslash \!\mathbf{0}}{}}\left[ \mathbf{g},\mathbf{g%
}^{\prime }\right] & =(g_{j}+k_{j})(g_{l}^{\prime }+k_{l})\hat{C}_{ijkl}(%
\mathbf{g}-\mathbf{g}^{\prime })-\omega ^{2}\hat{\rho}_{ik}(\mathbf{g}-%
\mathbf{g}^{\prime })  \notag \\
& \quad -\omega (g_{j}+k_{j})\hat{S}_{ijk}(\mathbf{g}-\mathbf{g}^{\prime
})-\omega (g_{j}^{\prime }+k_{j})\hat{S}_{{kji}}^{\ast }(%
\mathbf{g}^{\prime }-\mathbf{g}),
\end{align}%
where $\hat{S}_{ijk}\equiv \hat{S}_{\left( ij\right) k}=\hat{S}_{jik}$. Note
that since the matrix $\hat{\mathbf{G}}^{\overset{\backslash \!%
\mathbf{0}}{}}$ is Hermitian ($\hat{G}_{ip}^{\overset{\backslash \!\mathbf{0}%
}{}}\left[ \mathbf{g},\mathbf{g}^{\prime }\right] =\hat{G}_{pi}^{\overset{%
\backslash \!\mathbf{0}\ast }{}}\left[ \mathbf{g}^{\prime },\mathbf{g}\right]
$), the effective material tensors (\ref{-13}) established in this paper
recover the symmetries (\ref{-434}) of the corresponding tensors of the
initial periodic material (although not the sign definiteness, see \S \ref%
{4.1} for more details). The Willis formulation of elastodynamics
equations is therefore closed under the PWE homogenization.

\section{PWE effective parameters: derivation for a scalar wave model} \label{sec3}

Instead of proceeding with the elastodynamic equations \eqref{=64} in full,
we first provide a detailed derivation for the case of a scalar wave
equation. The scalar case is more revealing as it avoids the multitudinous
suffices of the equations of elasticity while retaining the essence of the
homogenization. We therefore ignore the obvious fact that an uncoupled
acoustic scalar wave in solids with $\mathbf{x}$-dependent material
parameters is restricted to $\mathbf{x\in \mathbb{R}}^{d}$ with $d=$1 or 2
since the scalar problem is purely a methodological shortcut. Once the
scalar derivation has been completed, it is straightforward to generalize
the results to the original elastodynamics system \eqref{=64} in $\mathbf{%
x\in \mathbb{R}}^{d},$ $d=$1, 2 or 3, see \S \ref{sec4}.

\subsection{Exact solution of the scalar wave problem}

For the scalar wave model, the displacement $u$, momentum $p,$ body force $f$
and density $\rho $ are scalars, while the stress $\mathbf{\sigma ,}$ strain 
${\pmb\varepsilon}=\mathbf{\nabla }u,$ strain source $\mathbf{\gamma }$,
and coupling parameter $\mathbf{S}$ are vectors and the 'shear' stiffness $%
\pmb{ \mu}$ is a $d\times d$ matrix in $\mathbf{\mathbb{C}}^{d}$
(disregarding the trivial case of $d=1$). According to (\ref{-434}), $\rho $
is real and $\pmb{ \mu}$ Hermitian while ${\pmb S}$ may be complex. Let
these material parameters be $\mathbf{T}$-periodic in the sense (\ref{19})
and take the wave variables and sources in Bloch form (\ref{2}). Then the
scalar-wave version of the governing equations (\ref{=64}) in component
form, with 
$_{,j}\equiv \partial /\partial x_{j}$, is 
\end{subequations}
\begin{equation}
-i\omega p=f+ik_{j}\sigma _{j}+\sigma _{j,j},\qquad 
\begin{pmatrix}
\sigma _{j} \\ 
\\ 
p%
\end{pmatrix}%
=%
\begin{pmatrix}
\mu _{jl} & S_{j} \\ 
&  \\ 
-S_{l}^{\ast } & \rho 
\end{pmatrix}%
\begin{pmatrix}
u_{,l}+ik_{l}u-\gamma _{l} \\ 
\\ 
-i\omega u%
\end{pmatrix}%
.  \label{4-1}
\end{equation}%
Applying Fourier transform (\ref{3-5}) reduces (\ref{4-1}) to the following
algebraic equations in $V_{\mathbf{g}}:$ 
\begin{equation}
\begin{aligned} -i\omega \hat{p}(\mathbf{g})&
=\hat{f}(\mathbf{g})+i(g_{j}+k_{j})\hat{\sigma}_{j}(\mathbf{g}), \\
\begin{pmatrix} \hat{\sigma}_{j}(\mathbf{g}) \\ \\
\hat{p}(\mathbf{g})\end{pmatrix}& =\sum_{\mathbf{g}^{\prime }\in \Gamma
}\begin{pmatrix} \hat{\mu}_{jl}(\mathbf{g}-\mathbf{g}^{\prime }) &
\hat{S}_{j}(\mathbf{g}-\mathbf{g}^{\prime }) \\ & \\ -\hat{S}_{l}^{\ast
}(\mathbf{g}-\mathbf{g}^{\prime }) &
\hat{\rho}(\mathbf{g}-\mathbf{g}^{\prime })\end{pmatrix}\begin{pmatrix}
i(k_{l}+g_{l}^{\prime })\hat{u}(\mathbf{g}^{\prime
})-\hat{\gamma}_{l}(\mathbf{g}^{\prime }) \\ \\ -i\omega
\hat{u}(\mathbf{g}^{\prime })\end{pmatrix} . \end{aligned}  \label{5}
\end{equation}%
It is convenient to pass to a matrix form in $V_{\mathbf{g}}$ while keeping
explicit the coordinate indices in $\mathbf{x}$-space (this makes the
eventual generalization to the fully elastodynamic problem clearer). Recall
the notation $\hat{\mathbf{h}}$ for the vector in $V_{\mathbf{g}}$ with the
components $\hat{h}(\mathbf{g})$ (see (\ref{3-5})), and introduce the
matrices $\hat{\mathbf{M}}_{jl},~\hat{\mathbf{D}}_{l},\hat{\mathbf{N}},~\hat{%
\mathbf{V}}_{l}$ ($\in V_{\mathbf{g}}\otimes V_{\mathbf{g}}$ when $j,l$ are
fixed) with the components 
\begin{equation}
\begin{aligned} \hat{M}_{jl}\left[ \mathbf{g,g}^{\prime }\right]
&={\hat{\mu}}_{jl}(\mathbf{g}-\mathbf{g}^{\prime }),\  \hat{D}_{l}\left[
\mathbf{g,g}^{\prime }\right] =(g_{l}+k_{l})\delta
_{\mathbf{g}\mathbf{g}^{\prime }}, \\ \hat{N}\left[ \mathbf{g,g}^{\prime
}\right] &={\hat{\rho}}(\mathbf{g}-\mathbf{g}^{\prime }),\ \ 
\hat{V}_{l}\left[ \mathbf{g,g}^{\prime }\right]
={\hat{S}}_{l}(\mathbf{g}-\mathbf{g}^{\prime }). \end{aligned}  \label{6a}
\end{equation}%
Note that $\hat{M}_{jl}\left[ \mathbf{g^{\prime },g}\right] =\hat{M}%
_{lj}^{\ast }\left[ \mathbf{g,g}^{\prime }\right] $ etc., i.e.\ the matrices $%
\hat{\mathbf{M}}_{jl},$~$\hat{\mathbf{D}}_{l}$ and\ $\hat{\mathbf{N}}$ are
Hermitian, while $\hat{\mathbf{V}}_{l}$ is not once ${S}_{l}\left( \mathbf{x}%
\right) $ is complex. Equations \eqref{5} can now be expressed in a matrix
form on $V_{\mathbf{g}}$ as 
\begin{equation}
\begin{aligned} -i\omega \hat{\mathbf p}& =\hat{\mathbf f}+i\hat{\pmb
D}_{j}\hat{\pmb\sigma }_{j}, \\ \begin{pmatrix} \hat{\pmb\sigma }_{j} \\ \\
\hat{\mathbf p}\end{pmatrix}& =\begin{pmatrix} \hat{\mathbf{M}}_{jl} &
\hat{\mathbf{V}}_{j} \\ & \\ -\hat{\mathbf{V}}_{l}^{+} &
\hat{\mathbf{N}}\end{pmatrix}\begin{pmatrix}
i\hat{\mathbf{D}}_{l}\hat{\mathbf{u}-\hat{\gamma}}_{l} \\ \\ -i\omega
\hat{\mathbf{u}}\end{pmatrix}=\begin{pmatrix} \hat{\mathbf{Q}}_j &
-\hat{\mathbf{M}}_{jl} \\ & \\ -\hat{\mathbf{R}} &
\hat{\mathbf{V}}_{l}^{+}\end{pmatrix}\begin{pmatrix} i\hat{\mathbf{u}} \\ \\
\hat{\mathbf{\gamma}}_{l}\end{pmatrix},\ \ \  \\ \mathrm{where\quad }&
\hat{\mathbf{Q}}_{j}=\hat{\mathbf{M}}_{jl}\hat{\mathbf{D}}_{l}-\omega
\hat{\mathbf{V}}_{j},\quad
\hat{\mathbf{R}}=\hat{\mathbf{V}}_{l}^{+}\hat{\mathbf{D}}_{l}+\omega
\hat{\mathbf{N}}. \end{aligned}  \label{7}
\end{equation}

Reduction to a single equation for the displacement provides a definition of
the impedance matrix $\hat{\mathbf{Z}}$ in the Fourier domain: 
\begin{equation}  \label{8}
\begin{aligned} \hat{\mathbf{Z}}\,\hat{\mathbf u} & =\hat{\mathbf
f}-i\hat{\mathbf Q}_{j}^{+}\hat{\pmb\gamma }_{j}, \quad \mathrm{where }\\
\hat{\mathbf{Z}}(\omega ,\mathbf{k})& \equiv
\hat{\mathbf{D}}_{j}\hat{\mathbf{M}}_{jl}\hat{\mathbf{D}}_{l}-\omega
(\hat{\mathbf{V}}_{j}^{+}\hat{\mathbf{D}}_{j}+\hat{\mathbf{D}}_{j}\hat{%
\mathbf{V}}_{j})-\omega ^{2}\hat{\mathbf{N}}\ \ 
\big(=\hat{\mathbf{D}}_{j}\hat{\mathbf{Q}}_{j}-\omega
\hat{\mathbf{R}}\big)=\hat{\mathbf{Z}}^{+}(\omega ,\mathbf{k}). \end{aligned}
\end{equation}
We note that the condition for existence of free waves in the absence of
sources, $\det \hat{\mathbf{Z}}(\omega ,\mathbf{k})=0,$ defines the Bloch
eigen-frequencies $\omega \left( \mathbf{k}\right) $. Our interest is
ultimately in homogenized equations of motion allowing for the Bloch waves.
It is however simpler to first proceed under the assumption that $\hat{%
\mathbf{Z}}(\omega ,\mathbf{k})$ is not singular ($\det \hat{\mathbf{Z}}\neq
0$). The issue of how to handle a singular $\hat{\mathbf{Z}}$ is 
resolved afterwards in \S \ref{reg}, by introduction of  a slightly
modified impedance.

Assuming $\hat{\mathbf{Z}}$ is non-singular the Green's matrix
function $\hat{\mathbf{G}}(\omega ,\mathbf{k})$ is defined as the inverse of $\hat{%
\mathbf{Z}}$.  The exact scalar solution then follows from \eqref{8} as 
\begin{equation}
\hat{\mathbf{u}}=\hat{\mathbf{G}}\hat{\mathbf{f}}-i\hat{\mathbf{G}}\hat{%
\mathbf{Q}}_{j}^{+}\hat{\pmb\gamma }_{j},\mathrm{\quad with\ }\hat{\mathbf{G}%
}=\hat{\mathbf{Z}}^{-1}\ \big(=\hat{\mathbf{G}}^{+}\big).  \label{9}
\end{equation}

\subsection{Homogenized equations} \label{3.2}

Taking the average (\ref{3-4}) of the wave solution \eqref{9} and the
constitutive equations \eqref{7}$_{2}$ yields 
\begin{equation}
\begin{aligned} \langle u\rangle & =\hat{\mathbf{e}}^{+} \hat{\mathbf{G}}
\hat{\mathbf{f}} -i\hat{\mathbf e}^{+}\hat{\mathbf G}\hat{\mathbf
Q}_{j}^{+}\hat{\pmb\gamma }_{j}, \\ \langle \sigma_{j}\rangle &
=i\hat{\mathbf{e}}^{+}\hat{\mathbf{Q}}_{j}\hat{\mathbf{G}} \hat{\mathbf{f}}
+\hat{\mathbf e}^{+}\big(\hat{\mathbf Q}_{j}\hat{\mathbf G}\hat{\mathbf
Q}_{l}^{+}-\hat{\mathbf{M}}_{jl}\big)\hat{\pmb\gamma }_{l}, \\ \langle
p\rangle & =-i\hat{\mathbf{e}}^{+}
\hat{\mathbf{R}}\hat{\mathbf{G}}\hat{\mathbf{f}} -\hat{\pmb
e}^{+}\big(\hat{\mathbf{R}}\hat{\mathbf G}\hat{\mathbf Q}_{l}^{+}-
\hat{\mathbf V}_{l}^{+}\big)\hat{\pmb\gamma }_{l}. \end{aligned}  \label{12}
\end{equation}%
Our objective is to rearrange eqs.\ \eqref{12} into a form   relating the
effective wave fields to the effective forcing terms with the latter viewed
as independent variables. As a first step in this direction, we insert the
effective sources (see eq.\ (\ref{3-2})) $\hat{\mathbf{\gamma }}_{j}^{%
\mathrm{eff}}=\left\langle \gamma _{j}\right\rangle \hat{\mathbf{e}}$ and $%
\hat{\mathbf{f}}^{\mathrm{eff}}=\left\langle f\right\rangle \hat{\mathbf{e}}$
in eqs.\ \eqref{12} and rewrite the result in a compact form 
\begin{equation}
\begin{aligned} \langle u\rangle & =a\left\langle f\right\rangle \,-i\beta
_{j}^{\ast }\,\left\langle \gamma _{j}\right\rangle , \\ \langle \sigma
_{j}\rangle & =i\beta _{j}\,\left\langle f\right\rangle
-A_{jl}\,\left\langle \gamma _{l}\right\rangle , \\ \langle p\rangle &
=-ib^{\ast }\,\left\langle f\right\rangle \,+\alpha _{j}^{\ast
}\,\left\langle \gamma _{j}\right\rangle , \end{aligned}  \label{121}
\end{equation}%
where the scalars $a$, $b\in \mathbb{C}$, the vectors ${\pmb\alpha }$, ${\pmb%
\beta }\in \mathbb{C}^{d}$ and the matrix $\mathbf{A}\in \mathbb{C}%
^{d}\otimes \mathbb{C}^{d}$ are 
\begin{equation}
\begin{aligned} & a=\hat{\mathbf{e}}^{+}
\hat{\mathbf{G}}\hat{\mathbf e},\quad
b=\hat{\mathbf e}^{+} \hat{\mathbf G}\hat{\mathbf{R}}^{+}\hat{\mathbf e},\quad
\beta _{j}=\hat{\mathbf e}^{+} \hat{\mathbf Q}_{j}\hat{\mathbf G}\hat{\mathbf e},
\\ & A_{jl}=\langle \mu _{jl}\rangle -
\hat{\mathbf{e}}^{+}\hat{\mathbf{Q}}_{j}\hat{\mathbf{G}}\hat{%
\mathbf{Q}}_{l}^{+}\hat{\mathbf{e}},\quad \alpha _{j}=\langle S_{j}\rangle
-\hat{\mathbf e}^{+}\hat{\mathbf Q}_{j}\hat{\mathbf
G}\hat{\mathbf{R}}^{+}\hat{\mathbf e} . \end{aligned}  \label{14}
\end{equation}%
Note 
that $a$ is real and $\mathbf{A}^{+}=\mathbf{A}$. Eliminating $\left\langle
f\right\rangle $ using \eqref{121}$_{1}$ allows us to recast 
eq.\ \eqref{121}$_{2}$ and \eqref{121}$_{3}$ in a form reminiscent of the Willis constitutive
relations, 
\begin{equation}
\begin{pmatrix}
\langle \sigma _{j}\rangle \\ 
\\ 
\langle p\rangle%
\end{pmatrix}%
=\frac{1}{a}%
\begin{pmatrix}
i\beta _{j} \\ 
\\ 
-ib^{\ast }%
\end{pmatrix}%
\langle u\rangle -%
\begin{pmatrix}
A_{jl}+\frac{1}{a}\beta _{j}\beta _{l}^{\ast } & X_{j} \\ 
&  \\ 
-\alpha _{j}^{\ast }-\frac{b^{\ast }}{a}\beta _{j}^{\ast } & Y%
\end{pmatrix}%
\begin{pmatrix}
\left\langle \gamma _{j}\right\rangle \\ 
\\ 
0%
\end{pmatrix}%
,  \label{-02}
\end{equation}%
where $X_{j}$ and $Y$ are at this stage arbitrary. Comparing (\ref{-02})
with the assumed structure of the effective constitutive relations applied
to fields in the presence of a strain source, eq.\ (\ref{00})$_{2}$ and
bearing in mind (\ref{3-2}) yields 
\begin{equation}
\begin{pmatrix}
\langle \sigma _{j}\rangle \\ 
\\ 
\langle p\rangle%
\end{pmatrix}%
=%
\begin{pmatrix}
\mu _{jl}^{\text{eff}} & S_{j}^{\text{eff}} \\ 
&  \\ 
-{S_{j}^{\text{eff}}}^{\ast } & \rho ^{\text{eff}}%
\end{pmatrix}%
\bigg\{%
\begin{pmatrix}
ik_{l} \\ 
\\ 
-i\omega%
\end{pmatrix}%
\langle u\rangle -%
\begin{pmatrix}
\left\langle \gamma _{j}\right\rangle \\ 
\\ 
0%
\end{pmatrix}%
\bigg\}.  \label{13}
\end{equation}%
Equating the contributions to $\langle \sigma _{j}\rangle $ and $\langle
p\rangle $ in \eqref{-02} and \eqref{13} from ${\pmb\gamma }$ gives,
in turn, 
\begin{equation}
{\mu }_{jl}^{\text{eff}}\ =A_{jl}+\frac{1}{a}{\beta }_{j}{\beta }_{l}^{\ast
},\qquad S_{j}^{\text{eff}}\ ={\alpha }_{j}+\frac{b}{a}{\beta }_{j}.
\label{17}
\end{equation}%
Full equivalence requires $Y=\rho ^{\text{eff}}$ and $\mathbf{X}=\mathbf{S}^{%
\text{eff}},$ which implies two more identities found by equating the
contributions from $\langle u\rangle $ to $\langle p\rangle $ and to $%
\langle \sigma _{j}\rangle $, respectively: 
\begin{equation}
\omega \rho ^{\text{eff}}\ =a^{-1}{b^{\ast }}-S{_{j}^{\text{eff}}}^{\ast
}k_{j},\qquad \omega S_{j}^{\text{eff}}\ ={\mu }_{jl}^{\text{eff}%
}k_{l}-a^{-1}{\beta }_{j}.  \label{18}
\end{equation}%
The first equation provides an expression for $\rho ^{\text{eff}}$ and the
second is an alternative to \eqref{17}$_{2}$ for $\mathbf{S}^{\text{eff}}\ $%
(equivalence of \eqref{17}$_{2}$ and \eqref{18}$_{2}$ is noted below).

Combining eqs.\  \eqref{17} and \eqref{18}$_1$ and reinstating the notations
from (\ref{14}) yields the effective parameters in the form 
\begin{subequations}
\label{20}
\begin{align}
\mu _{jl}^{\text{eff}}(\omega ,\mathbf{k})& =\langle \mu _{jl}\rangle -\hat{%
\mathbf{e}}^{+}\hat{\mathbf{Q}}_{j}\hat{\mathbf{G}}\hat{\mathbf{Q}}_{l}^{+}%
\hat{\mathbf{e}}+\frac{(\hat{\mathbf{e}}^{+}\hat{\mathbf{Q}}_{j}\hat{\mathbf{%
G}}\hat{\mathbf{e}})\, ( \hat{\mathbf{e}}^{+}\hat{\mathbf{G}}\hat{\mathbf{Q}}%
_{l}^{+}\hat{\mathbf{e}} ) ^{+}}{\hat{\mathbf{e}}^{+}\hat{\mathbf{G}}\hat{%
\mathbf{e}}},  \label{20b} \\
\rho ^{\text{eff}}\ (\omega ,\mathbf{k})& =\langle \rho \rangle +\mathbf{%
\hat{e}}^{+}\hat{\mathbf{R}}\hat{\mathbf{G}}{\hat{\mathbf{R}}}^{+}\hat{%
\mathbf{e}}-\frac{|\hat{\mathbf{e}}^{+}\hat{\mathbf{G}}{\hat{\mathbf{R}}}^{+}%
\mathbf{\hat{e}}|^{2}}{\hat{\mathbf{e}}^{+}\hat{\mathbf{G}}\hat{\mathbf{e}}},
\label{20a} \\
S_{j}^{\text{eff}}(\omega ,\mathbf{k})& =\langle S_{j}\rangle -\hat{\mathbf{e}}^{+}\hat{\mathbf{Q}}_{j}\hat{\mathbf{G}}{\hat{\mathbf{R}}^{+}\hat{\mathbf{e}}}+ 
\frac{( \hat{\mathbf{e}}^{+}\hat{\mathbf{G}}{\hat{\mathbf{R}}^{+}\hat{\mathbf e}}) ( 
\hat{\mathbf{e}}^{+}\hat{\mathbf{G}}\hat{\mathbf{Q}}_{j}^{+}\mathbf{\hat{\mathbf e}}
) }{\hat{\mathbf{e}}^{+}\hat{\mathbf{G}}\hat{\mathbf{e}}}.  \label{20c}
\end{align}%
In the derivation of (\ref{20}), we have taken note that $\hat{\mathbf{Z}}=%
\hat{\mathbf{D}}_{j}\hat{\mathbf{Q}}_{j}-\omega \hat{\mathbf{R}}=\hat{%
\mathbf{Z}}^{+}$ and used identities such as 
\end{subequations}
\begin{equation}
\mathbf{k}\cdot {\pmb \beta}^{\ast } =k_{j}\hat{\mathbf{e}}^{+}\mathbf{%
\hat{G}\hat{Q}}_{j}^{+}\hat{\mathbf{e}} =\hat{\mathbf{e}}^{+}\hat{\mathbf{G}}%
\hat{\mathbf{Q}}_{j}^{+}\hat{\mathbf{D}}_{j}\hat{\mathbf{e}}=\hat{\mathbf{e}}%
^{+}\hat{\mathbf{G}}\left( \hat{\mathbf{Z}}+\omega \hat{\mathbf{R}}%
^{+}\right) \hat{\mathbf{e}}=1+\omega \hat{\mathbf{e}}^{+}\hat{\mathbf{G}}{%
\hat{\mathbf{R}}^{+}\hat{\mathbf{e}}}.  \label{21}
\end{equation}%
The equivalence of the expressions \eqref{17}$_2$ and \eqref{18}$_2$ for $%
\mathbf{S}^{\text{eff}}$ follows from using the explicit formulas \eqref{20}
and  identities of the above type.

\subsection{Regularized form of the effective parameters}

\label{reg}

By definition (\ref{9}), the Green's function $\mathbf{\hat{G}}\left( \omega
,\mathbf{k}\right) =\mathbf{\hat{Z}}^{-1}\left( \omega ,\mathbf{k}\right) $
diverges on the Bloch branches $\Omega _{\mathrm{B}}$ (3-parameter
manifold in $\left\{ \omega ,\mathbf{k}\right\} $-space) defined by the
Bloch dispersion equation 
\begin{equation}
\Omega _{\mathrm{B}}\ni \left( \omega ,\mathbf{k}\right) _{\mathrm{B}}:\
\det \mathbf{\hat{Z}}=0\ \Leftrightarrow \ \mathbf{\hat{Z}\mathbf{\hat{u}}}_{%
\mathrm{B}}=\mathbf{0},  \label{B}
\end{equation}%
where the null vector $\mathbf{\hat{u}}_{\mathrm{B}}$ of $\mathbf{\hat{Z}}$
is the Bloch eigen-mode in the Fourier domain. At the same time, the
effective dynamic parameters given by eqs.\ \eqref{20} generally remain
finite on $\Omega _{\mathrm{B}}$ (accidental exceptions such as degenerate
and certain other points on $\Omega _{\mathrm{B}}$ are disregarded for the
moment and will be discussed in \S \ref{excep}). This can be shown by
inspecting the limit of eqs.\ \eqref{20} as $\omega ,\mathbf{k}$ tend to a
(single) point $\left( \omega ,\mathbf{k}\right) _{\mathrm{B}}$ on a Bloch
branch, from which it is seen that the right-hand side members of \eqref{20}
that diverge on $\Omega _{\mathrm{B}}$ in fact compensate and cancel each
other. However, such limiting behavior of eqs.\ \eqref{20} prevents their
numerical implementation by evaluating each member independently before
summing them up. The difficulty can be circumvented if we recast eqs.\ %
\eqref{20} in the analytically equivalent but explicitly different form that
is rid of the terms diverging on $\Omega _{\mathrm{B}}.$ Since the
divergence in\ \eqref{20} is due to Green's function $\mathbf{\hat{G},}$ the
idea is to extract the part $\hat{\mathbf{G}}_{\mathrm{reg}}$ of $\mathbf{%
\hat{G}}$ which is regular on $\Omega _{\mathrm{B}}.$ Consider such a
partitioning of $\mathbf{\hat{G}}$ in the general form $\mathbf{\hat{G}}%
\left( \omega ,\mathbf{k}\right) =\hat{\mathbf{G}}_{\mathrm{reg}}+\epsilon
^{-1}\mathbf{\hat{v}\hat{v}}^{+},$ where $\mathbf{\hat{v}\rightarrow \hat{u}}%
_{\mathrm{B}}$ and $\epsilon \rightarrow 0$ as $\omega ,\mathbf{k}%
\rightarrow \left( \omega ,\mathbf{k}\right) _{\mathrm{B}}.$ It is not
unique. For instance, one standard way is to invoke the spectral
decomposition 
\begin{equation}
\hat{\mathbf{Z}}\left( \omega ,\mathbf{k}\right) =\hat{\mathbf{Z}}^{\prime
}+\varsigma \mathbf{\hat{v}\hat{v}}^{+},\ \hat{\mathbf{G}}\left( \omega ,%
\mathbf{k}\right) =\hat{\mathbf{G}}^{\prime }+\varsigma ^{-1}\mathbf{\hat{v}%
\hat{v}}^{+},  \label{psi}
\end{equation}%
where $\varsigma $ and $\mathbf{\hat{v}}$ are the corresponding eigenvalue
and eigenvector of $\hat{\mathbf{Z}}\,(=\hat{\mathbf{Z}}^{+})$ which satisfy 
$\mathbf{\hat{v}\rightarrow \hat{u}}_{\mathrm{B}}$ and $\varsigma
\rightarrow 0$ as $\omega ,\mathbf{k}\rightarrow \left( \omega ,\mathbf{k}%
\right) _{\mathrm{B}},$ and $\hat{\mathbf{G}}^{\prime }=\hat{\mathbf{Z}}%
^{\prime ^{-1}}$ (the pseudoinverse of $\hat{\mathbf{Z}}$) is regular on $%
\Omega _{\mathrm{B}}.$ Substituting (\ref{psi})$_{2}$ for $\mathbf{\hat{G}}$
in \eqref{20} removes the diverging terms and yields an explicit form
expressed through $\hat{\mathbf{G}}^{\prime }$. Unfortunately, explicit
calculation of $\hat{\mathbf{G}}^{\prime }$ is relatively laborious. In this
regard, we advocate the use of another partitioning of $\hat{\mathbf{G}}$
into regular and diverging parts which requires merely singling out Fourier
components with zero and non-zero $\mathbf{g}$. This is an essential
ingredient of our homogenization method that ensures both robust and
straightforward numerical implementation.

With the above idea in mind, denote by $V_{\mathbf{g}\neq \mathbf{0}}$ the
vector space that is associated with the set ${\Gamma \backslash \mathbf{0},}
$ i.e.\ is orthogonal to the vector $\mathbf{\hat{e}}$ (see (\ref{101}) and (%
\ref{3-4})). Let $\hat{\mathbf{Z}}_{\backslash \mathbf{0}}$ be the
restriction of $\hat{\mathbf{Z}}$ over this space $V_{\mathbf{g}\neq \mathbf{0}}$
and $\hat{\mathbf{G}}_{\backslash \mathbf{0}}$ be the inverse of $\hat{%
\mathbf{Z}}_{\backslash \mathbf{0}}$ on $V_{\mathbf{g}\neq \mathbf{0}}:$%
\begin{equation}
\hat{\mathbf{Z}}_{\backslash \mathbf{0}}\equiv \left. \hat{\mathbf{Z}}%
\right\vert _{\Gamma \backslash \mathbf{0}};\quad \hat{\mathbf{G}}%
_{\backslash \mathbf{0}}\equiv \hat{\mathbf{Z}}_{\backslash \mathbf{0}%
}^{-1}\ \ (\mathrm{for}\ \det \hat{\mathbf{Z}}_{\backslash \mathbf{0}}\neq 0)
\label{-4-}
\end{equation}%
(note that $\hat{\mathbf{G}}_{\backslash \mathbf{0}}$ as defined is \textit{%
not} a restriction of $\hat{\mathbf{G}}=\hat{\mathbf{Z}}^{-1}$ on $V_{%
\mathbf{g}\neq \mathbf{0}}$). Introduce the vectors $\mathbf{\hat{w}}$, $%
\mathbf{\hat{q}}_{j}$, $\mathbf{\hat{r}}$ $\in V_{\mathbf{g}\neq \mathbf{0}}$%
, 
\begin{equation}
\begin{aligned} 
& \qquad \qquad 
\mathbf{\hat{w}} \equiv k_{j}\mathbf{\hat{q}}_{j}-\omega \mathbf{\hat{r}} 
\quad \text{where} \qquad
\\ 
& \begin{array}{c} 
\mathbf{\hat{q}}_{j} \equiv \mathbf{\hat{Q}}_{j}^{+}\mathbf{\hat{\mathbf{e}}-(
\hat{\mathbf{e}}\hat{Q}}_{j}^{+}\mathbf{\hat{\mathbf{e}} ) \hat{\mathbf{e}} },
\\ 
\mathbf{\hat{r}} \equiv \mathbf{\hat{R}}^{+}\mathbf{\hat{e}-} (
\mathbf{\hat{\mathbf{e}}}^{+}\mathbf{\hat{R}}^{+}\mathbf{\hat{\mathbf{e}}} )
\mathbf{\hat{\mathbf{e}} },
\end{array}
 \Rightarrow 
\begin{array}{c} 
\hat{q}_{j}\left( \mathbf{g}\right)
=(g_{l}+k_{l})\hat{\mu}_{jl}\left( \mathbf{g}\right) -\omega
\hat{S}_{j}^{\ast }\left( \mathbf{g}\right) , 
\\
\hat{r}\left(
\mathbf{g}\right) =\omega \hat{\rho}\left( \mathbf{g}\right)
+(g_{j}+k_{j})\hat{S}_{j}\left( \mathbf{g}\right) ,
\end{array}
\mathrm{for}\ 
\mathbf{g}\in \Gamma \backslash \mathbf{0} . 
\end{aligned}
\label{-81}
\end{equation} 
Also denote $\langle \hat{Z}\rangle =\langle \hat{\mathbf{Z}}\mathbf{\hat{e}}%
\rangle $ and $\langle \hat{G}\rangle =\langle \hat{\mathbf{G}}\mathbf{\hat{e%
}}\rangle $ which is a slight generalization of the average \eqref{3-4} to
include the projection of a matrix onto $\mathbf{g},\,\mathbf{g}\prime =%
\mathbf{0}$.  Thus  
\begin{equation}
\begin{aligned} 
\langle\hat{Z} \rangle &\equiv \hat{Z}
\left[\mathbf{0,0}\right]
={\mathbf{\hat{e}}^{+}\mathbf{\hat{Z}\hat{e}~}}={k_{j}k_{l}\langle \mu
_{jl}\rangle -\omega ^{2}\langle \rho \rangle -\omega k_{j}\langle
S_{j}+S_{j}^{\ast }\rangle ,\ } 
\\ \langle\hat{G} \rangle &\equiv
{{\hat{G}}}\left[ \mathbf{0,0}\right]
=\mathbf{\hat{e}}^{+} \mathbf{\hat{G}\hat{e}}
=\det {\mathbf{\hat{Z}}}_{\backslash \mathbf{0}}\det {\mathbf{\hat{Z}}}^{-1}.
\end{aligned}  \label{82}
\end{equation}
Using the notations (\ref{-81}) and (\ref{82}), we note the identical
decompositions   
\begin{equation}
{\mathbf{\hat{Z}}}\left( \omega ,\mathbf{k}\right) ={\mathbf{\hat{Z}}}%
_{\backslash \mathbf{0}}+\mathbf{\hat{w}\hat{e}}^{+}+\mathbf{\hat{e}\hat{w}}%
^{+}+\langle \hat{Z}\rangle \mathbf{\hat{e}\hat{e}}^{+},\quad \det {\mathbf{%
\hat{Z}}}=\langle \hat{Z}\rangle \det {\mathbf{\hat{Z}}}_{\backslash \mathbf{%
0}}-\mathbf{w}^{+}\overline{\mathbf{\hat{Z}}}_{\backslash \mathbf{0}}\mathbf{%
w,}  \label{83}
\end{equation}%
where $\overline{\mathbf{\hat{Z}}}_{\backslash \mathbf{0}}$ is adjugate of $%
\mathbf{\hat{Z}}_{\backslash \mathbf{0}}$. Hence if $\det {\mathbf{\hat{Z}}}%
_{\backslash \mathbf{0}}\neq 0$ as assumed in (\ref{-4-})$_{2}$, then 
\begin{subequations}
\label{84}
\begin{align}
\langle \hat{G}\rangle ^{-1}& =\langle \hat{Z}\rangle -{\mathbf{\hat{w}}^{+}%
\mathbf{\hat{G}_{\backslash \mathbf{0}}\hat{w}}},  \label{84aa} \\
\det {\mathbf{\hat{Z}}}\left( \omega ,\mathbf{k}\right) & =\langle \hat{G}%
\rangle ^{-1}\det {\mathbf{\hat{Z}}}_{\backslash \mathbf{0}}{\mathbf{~}}%
\left( =0\ \Leftrightarrow \langle \hat{G}\rangle ^{-1}=0\ \mathrm{on\ }%
\Omega _{\mathrm{B}}\right) ,  \label{84a} \\
{\mathbf{\hat{Z}}}\mathbf{\hat{v}}& =\langle \hat{G}\rangle ^{-1}\mathbf{%
\hat{e}\quad }\mathrm{with\ }\mathbf{\hat{v}}\left( \omega ,\mathbf{k}%
\right) =\mathbf{\hat{e}}-{\mathbf{\hat{G}}}_{\backslash \mathbf{0}}{\mathbf{%
{\hat{w}}}}\ \left( \rightarrow \mathbf{\hat{u}}_{\mathrm{B}}\ \mathrm{as\ }%
\omega ,\mathbf{k}\rightarrow \Omega _{\mathrm{B}}\right) ,  \label{84b} \\
{\mathbf{\hat{G}}}\left( \omega ,\mathbf{k}\right) & ={\mathbf{\hat{G}}}%
_{\backslash \mathbf{0}}+\langle \hat{G}\rangle \,\mathbf{\hat{v}\hat{v}}%
^{+},  \label{84c}
\end{align}
\end{subequations}
where eq.\ (\ref{84c}) yields the sought-for partitioning ${\mathbf{\hat{G}}}
$ into two parts, one regular and the other diverging on Bloch branches $%
\Omega _{\mathrm{B}}$. Note that generally (i.e.\ for ${\mathbf{\hat{G}}}%
_{\backslash \mathbf{0}}{\mathbf{{\hat{w}\neq 0}}}$) the quantities ${%
\mathbf{\hat{G}}}_{\backslash \mathbf{0}}$ $\,$and $\mathbf{\hat{v}~}\left(
\rightarrow \mathbf{\hat{u}}_{\mathrm{B}}\right) $ in (\ref{84c}) are
different from, respectively, the pseudoinverse ${\mathbf{\hat{G}}}^{\prime
} $ and the eigenvector $\mathbf{\hat{v}~}\left( \rightarrow \mathbf{\hat{u}}%
_{\mathrm{B}}\right) $ in (\ref{psi}).

Writing  (\ref{84c}) in the transparent form 
\begin{equation}  \label{843}
{\mathbf{\hat{G}}}_{\backslash \mathbf{0}}
= {\mathbf{\hat{G}}} - \big(  \mathbf{\hat{e}}^+  \mathbf{\hat{G}}\mathbf{\hat{e}}\big)^{-1}
\, 
\mathbf{\hat{G}}\mathbf{\hat{e}} \mathbf{\hat{e}}^+  \mathbf{\hat{G}}
\end{equation}
and using the notations (\ref{-81}), reduces eqs.\  \eqref{20} to the simpler form 
\begin{subequations}
\label{013}
\begin{align}
\mu _{jl}^{\text{eff}}(\omega ,\mathbf{k})& =\langle \mu _{jl}\rangle -\hat{%
\mathbf{q}}_{j}^{+}\hat{\mathbf{G}}_{\backslash \mathbf{0}}\mathbf{\hat{q}}%
_{l},  \label{013a} \\
S_{j}^{\text{eff}}(\omega ,\mathbf{k})& =\langle S_{j}\rangle -\hat{\mathbf{q%
}}_{j}^{+}\hat{\mathbf{G}}_{\backslash \mathbf{0}}\mathbf{\hat{r}},
\label{013b} \\
\rho ^{\text{eff}}\ (\omega ,\mathbf{k})& =\langle \rho \rangle +\hat{%
\mathbf{r}}^{+}\hat{\mathbf{G}}_{\backslash \mathbf{0}}\mathbf{\hat{r}.}
\label{013c}
\end{align}%
It is emphasized that the effective parameters explicitly defined in (\ref{013}) are by
construction the same as those in \eqref{20}. At the same time the terms in %
\eqref{20} divergent on $\Omega _{\mathrm{B}}$ are eliminated in (\ref{013}), 
in this sense (\ref{013}) are {uniformly convergent}. The new expressions (%
\ref{013}) are determined from the regular part ${\mathbf{\hat{G}}}%
_{\backslash \mathbf{0}}$ of the Green's function associated with $\mathbf{g}%
\in \Gamma \backslash \mathbf{0}$ which is straightforward to compute;
moreover, their explicit form is simpler than that of \eqref{20}.

\subsection{Exceptional cases}   \label{excep}

The expressions for the effective parameters (\ref{013}) defined by the
matrix ${\mathbf{\hat{G}}}_{\backslash \mathbf{0}}={\mathbf{\hat{Z}}}%
_{\backslash \mathbf{0}}^{-1}$ are obtained for the general case of
non-singular $\mathbf{\hat{Z}}_{\backslash \mathbf{0}}$. 
If ${\mathbf{\hat{Z}}}_{\backslash \mathbf{0}}$ happens to be singular, then
numerical implementation of (\ref{013})  produces diverging terms (though
eqs.\ (\ref{20}) expressed via ${\mathbf{\hat{G}=\hat{Z}}}^{-1}$ remain
finite for $\det {\mathbf{\hat{Z}}}_{\backslash \mathbf{0}}=0$ so long as $%
\det {\mathbf{\hat{Z}}}\neq 0$). The properties of the wave solutions that occur specifically at values of $%
\omega ,\mathbf{k}$ rendering ${\mathbf{\hat{Z}}}_{\backslash \mathbf{0}%
}\left( \omega ,\mathbf{k}\right) $ singular are of interest. Denote the set of these
exceptional points (3-parameter manifold in $\left\{ \omega ,\mathbf{k}%
\right\} $-space) by $\Omega _{\mathrm{exc}}$ so that 
\end{subequations}
\begin{equation}
\Omega _{\mathrm{exc}}\ni \left( \omega ,\mathbf{k}\right) _{\mathrm{exc}}:\
\det \mathbf{\hat{Z}}_{\backslash \mathbf{0}}=0.  \label{ex1}
\end{equation}%
Note that eqs.\  (\ref{84}) are invalid on $\Omega _{\mathrm{exc}}.$

Consider first 'generic' points of $\Omega _{\mathrm{exc}}$ which lie off
the Bloch branches (\ref{B}), i.e.\ $\left( \omega ,\mathbf{k}\right) _{%
\mathrm{exc}}\notin \Omega _{\mathrm{B}}$ ($\Rightarrow \det {\mathbf{\hat{Z}%
}}\neq 0$)$.$ Then by \eqref{82}$_{2}$ $\langle \hat{G}\rangle =\langle {%
\mathbf{\hat{G}\hat{e}}}\rangle =0$ and hence the averaged wave (\ref{121})$%
_{1}$ is $\left\langle u\right\rangle =i\langle \hat{Z}\rangle ^{-1}\gamma
_{j}\mathbf{\hat{w}}^{+}\overline{\mathbf{\hat{Z}}}_{\backslash \mathbf{0}}%
\mathbf{\hat{q}}_{j}.$ It is generally nonzero, which is not surprising as $%
\left\langle u\right\rangle =0$ would correspond to the zero effective
response to the non-zero effective forcing. At the same time, the wave
response to a pure force source ($\gamma _{j}=0$) at the points $\left(
\omega ,\mathbf{k}\right) _{\mathrm{exc}}\notin \Omega _{\mathrm{B}}$ has a
zero mean over the unit cell $\mathbf{T}$ in $\mathbf{x}$-space. It is
therefore somewhat satisfying and physically sensible that the effective
parameters (\ref{013}) indicate such exceptional points by becoming
numerically divergent.

Now assume a generally non-empty set $\Omega _{\mathrm{exc}}\cap \Omega _{%
\mathrm{B}}$ (2-parameter manifold in $\left\{ \omega ,\mathbf{k}\right\} $%
-space) of the points of Bloch branches (\ref{B}) at which ${\mathbf{\hat{Z}}%
}_{\backslash \mathbf{0}}$ happens to be singular. By (\ref{83})$_{2}$,
simultaneous equalities $\det {\mathbf{\hat{Z}}}=0$ and $\det {\mathbf{\hat{Z%
}}}_{\backslash \mathbf{0}}=0$ yield $\mathbf{w}^{+}\overline{\mathbf{\hat{Z}%
}}_{\backslash \mathbf{0}}\mathbf{w}=0.$ It is natural to assume that $%
\overline{\mathbf{\hat{Z}}}_{\backslash \mathbf{0}}\mathbf{w}$ is nonzero.
Then it is a null vector of ${\mathbf{\hat{Z}}}_{\backslash \mathbf{0}}$ and
it is orthogonal to $\mathbf{w}$. Hence, by (\ref{83})$_{1}$, $\overline{%
\mathbf{\hat{Z}}}_{\backslash \mathbf{0}}\mathbf{w}$ is also a null vector
of ${\mathbf{\hat{Z},}}$ i.e.\ it is a Bloch mode $\mathbf{\hat{u}}_{\mathrm{B%
}}$. Thus, if $\det {\mathbf{\hat{Z}}}_{\backslash \mathbf{0}}=0$
incidentally occurs at some point of a Bloch branch, then the Bloch mode at
this point is $\mathbf{\hat{u}}_{\mathrm{B}}=\overline{\mathbf{\hat{Z}}}%
_{\backslash \mathbf{0}}\mathbf{w}$, which is by construction orthogonal to $%
\mathbf{e}$ and so its original $u_{\mathrm{B}}\left( \mathbf{x}\right) $ in 
$\mathbf{x}$-space has zero mean $\left\langle u_{\mathrm{B}}\right\rangle
=0 $ over the unit cell $\mathbf{T}$. Let us call it a \textit{zero-mean
Bloch mode}. This is in contrast to 'normal' points of $\Omega _{\mathrm{B}}$
where the Bloch modes are defined by (\ref{84b}) as $\mathbf{\hat{u}}_{%
\mathrm{B}}=\mathbf{\hat{e}}-{\mathbf{\hat{G}}}_{\backslash \mathbf{0}}{%
\mathbf{{\hat{w}}}}$ and thus have non-zero mean $\left\langle u_{\mathrm{B}%
}\right\rangle =1.$ Since zero-mean Bloch modes are zero (trivial) solutions
of the effective equations for free waves, their explicit exclusion from the
framework of the homogenization theory, as signalled by divergence of the
effective material parameters (\ref{013}), is physically consistent. Note
that the divergence at the points of zero-mean Bloch modes is 'genuine'
(analytical), i.e.\ it occurs regardless of which explicit expressions are
used for the effective parameters (unless more restrictions are imposed 
apart from $\Omega _{\mathrm{exc}}\cap \Omega _{\mathrm{B}}$, such as $%
\overline{\mathbf{\hat{Z}}}_{\backslash \mathbf{0}}\mathbf{w=0}$).

{In conclusion, we note another exceptional case for the effective parameters
(\ref{013}) is the possible existence of double points (self-intersections)
on Bloch branches $\Omega _{\mathrm{B}}$. It can be shown, using e.g.\ the
spectral decomposition and the pseudoinverse of ${\mathbf{\hat{Z}}}$, that a
double point on $\Omega _{\mathrm{B}}$ causes a genuine (in the sense
mentioned above) divergence of the effective parameters, which conforms to
the fact that a double point admits the construction of a zero-mean Bloch
mode $\left\langle u_{\mathrm{B}}\right\rangle =0$ from a pair of null
vectors of ${\mathbf{\hat{Z}}}$.
}

\subsection{Summary for a scalar wave problem}

The equations governing the averaged effective fields $u^{\text{eff}}$, ${%
\pmb\sigma }^{\text{eff}}$, $p^{\text{eff}}$ and ${\pmb\varepsilon}^{\text{%
eff}}\ =\mathbf{\nabla }u^{\text{eff}}$ and their relation to the averaged
applied body force $f^{\text{eff}}$ and strain source ${\pmb\gamma }^{\text{%
eff}}$ (see (\ref{3-2})) are of Willis form: 
\begin{equation}
\dot{p}^{\text{eff}}\ =\text{div}\mathbf{\sigma }^{\text{eff}}\ +f^{\text{eff%
}},\qquad 
\begin{pmatrix}
\mathbf{\sigma }^{\text{eff}} \\ 
\\ 
p^{\text{eff}}%
\end{pmatrix}%
=%
\begin{pmatrix}
{\pmb\mu }^{\text{eff}} & {\mathbf{S}}^{\text{eff}} \\ 
&  \\ 
-{\ {\mathbf{S}}^{\text{eff}}}^{+} & \rho ^{\text{eff}}%
\end{pmatrix}%
\begin{pmatrix}
{\pmb\epsilon }^{\text{eff}}\ -{\pmb\gamma }^{\text{eff}} \\ 
\\ 
\dot{u}^{\text{eff}}%
\end{pmatrix}%
,  \label{-0}
\end{equation}%
where $\rho ^{\text{eff}}$, $\mu _{jl}^{\text{eff}}$ and $S_{j}^{\text{eff}}$
are non-local in space and time. They are defined in the transform domain by
eqs.\ (\ref{013}) with the following explicit form 
\begin{subequations}
\label{03}
\begin{align}
& \left. 
\begin{array}{c}
\mu _{jl}^{\text{eff}}(\omega ,\mathbf{k}) \\ 
\\ 
{S_{j}^{\text{eff}}}(\omega ,\mathbf{k}) \\ 
\\ 
\rho ^{\text{eff}}\ (\omega ,\mathbf{k})%
\end{array}%
\right\} =\left\{ 
\begin{array}{c}
\langle \mu _{jl}\rangle \\ 
\\ 
\langle S_{j}\rangle \\ 
\\ 
\langle \rho \rangle%
\end{array}%
\right.  \notag \\
& +%
\begin{matrix}
\sum\limits_{\mathbf{g},\mathbf{g}^{\prime }\in \Gamma \backslash \mathbf{0}}%
\end{matrix}%
\hat{G}_{\backslash \mathbf{0}}\left[ \mathbf{g},\mathbf{g}^{\prime }\right]
\times \left\{ 
\begin{array}{l}
-\left[ \hat{\mu}_{jn}(-\mathbf{g})(k_{n}+g_{n})-\omega \hat{S}_{j}(-\mathbf{g}%
)\right] \left[ \hat{\mu}_{lp}(\mathbf{g}^{\prime })(k_{p}+g_{p}^{\prime
})-\omega \hat{S}_{l}(\mathbf{g}^{\prime })\right] , \\ 
\\ 
-\left[ \hat{\mu}_{jl}(-\mathbf{g})(k_{l}+g_{l})-\omega \hat{S}_{j}(-\mathbf{g}%
)\right] \left[ \omega \hat{\rho}(\mathbf{g}^{\prime
})+(k_{n}+g_{n}^{\prime })\hat{S}_{n}(\mathbf{g}^{\prime } )\right] , \\ 
\\ 
\left[ \omega \hat{\rho}(-\mathbf{g})+(k_{j}+g_{j})\hat{S}_{j}(-\mathbf{g})%
\right] \left[ \omega \hat{\rho}(\mathbf{g}^{\prime })+(k_{l}+g_{l}^{\prime
})\hat{S}_{l}(\mathbf{g}^{\prime })\right] ,%
\end{array}%
\right.  \label{03a}
\end{align}%
where $\mathbf{\hat{G}}_{\backslash \mathbf{0}}=\mathbf{\hat{G}}_{\backslash 
\mathbf{0}}^{+}$ is the inverse to the matrix $\mathbf{\hat{Z}}_{\backslash 
\mathbf{0}}$ with the components 
\begin{align}
\hat{Z}_{\backslash \mathbf{0}}\left[ \mathbf{g},\mathbf{g}^{\prime }\right]
& =\bigg( (k_{j}+g_{j})(k_{l}+g_{l}^{\prime })\hat{\mu}_{jl}(\mathbf{g}-%
\mathbf{g}^{\prime })-\omega ^{2}\hat{\rho}(\mathbf{g}-\mathbf{g}^{\prime })
\notag \\
& \qquad \quad \left. -\omega (k_{j}+g_{j})\hat{S}_{j}(\mathbf{g}-\mathbf{g}%
^{\prime })-\omega (k_{j}+g_{j}^{\prime })\hat{S}_{j}^{\ast }(\mathbf{g}^{\prime } - 
\mathbf{g} )\bigg) \right|_{\mathbf{g},\mathbf{g}^{\prime }\neq 0}.
\label{03b}
\end{align}
Note that $\mathbf{\mu }^{\text{eff}}=\mathbf{\mu }^{\text{eff}^{+}} $,  $
\rho ^{\text{eff}}$ is real and  ${S_{j}^{\text{eff}}}$ is complex-valued, see 
\S \ref{4.1}.

The effective parameters defined by eqs.\ (\ref{03}) are regular at any $%
\omega ,~\mathbf{k}$ excluding possible double points (intersections) of
Bloch branches, and the exceptional points $\left( \omega ,\mathbf{k}\right)
_{\mathrm{exc}}\in \Omega _{\mathrm{exc}}$ where $\det \mathbf{\hat{Z}}%
_{\backslash \mathbf{0}}=0$ (a numerical divergence  appears in the
vicinity of such points). If a point $\left( \omega ,\mathbf{k}\right) _{%
\mathrm{exc}}$ incidentally occurs on a Bloch branch then the corresponding
Bloch eigen-mode $u_{\mathrm{B}}\left( \mathbf{x}\right) $ has a zero mean $%
\left\langle u_{\mathrm{B}}\right\rangle =0$ over the unit cell $\mathbf{T}$%
. Zero-mean Bloch mode can also exist at the double point of a Bloch branch.
Barring these exceptions, eqs.\ (\ref{03}) are regular on Bloch branches
which are defined by the dispersion equation (see \eqref{84aa}) 
\end{subequations}
\begin{equation}
\langle \hat{G}\rangle ^{-1}\left( \omega ,\mathbf{k}\right) =0\ \ \big(\det 
\mathbf{\hat{Z}}_{\backslash \mathbf{0}}\neq 0\big).  \label{04}
\end{equation}%
The term $\langle \hat{G}\rangle ^{-1}\equiv Z^{\text{eff}}$ which defines
the dispersion relation is expressed as a more physically meaningful
quantity in \S \ref{imp}.

\section{Elastodynamic effective parameters}

\label{sec4}

The procedure of adapting the scalar acoustic model to the vector
elastodynamic model requires that the scalar and vector variables $u$, $p$, $%
f$ and ${\pmb\sigma }$, ${\pmb\varepsilon}$, ${\pmb\gamma }$ of the
acoustic case become, respectively, vectors and symmetric second order
tensors in $\mathbb{\mathbb{C}}^{3}$. Accordingly, the scalar density $\rho
, $ coupling vector ${\mathbf{S}}$ and the stiffness matrix ${\pmb\mu }$
increase by two orders as tensor and become   $\mathbf{%
\rho }$, $\mathbf{S}$ and $\mathbf{C}$ defined in \S \ref{2.1}. The
component form of the various quantities for the two problems are related as
follows: 
\begin{align}
\text{scalar system}\qquad \qquad &  \quad \qquad \qquad \text{%
elastodynamic problem}  \notag  \label{-122} \\
u,\,p,\,f,\,\sigma _{j},\ \varepsilon _{j},\ \gamma _{j},\,\rho ,\
S_{j},\,\mu _{jl}\quad & \rightarrow \quad u_{i},\,p_{i},\,f_{i},\,\sigma
_{ij},\ \varepsilon _{ij},\ \gamma _{ij},\,\rho _{ik},\ S_{ijk},\,C_{ijkl}.
\end{align}%
Based on this equivalence the generalization of the scalar eqs.\ (\ref{013})
proceeds by first replacing the single suffix $j$ in the former with $ij$ in
the latter, then using definitions such as those in eq.\ \eqref{-81} to
assign additional suffices, ultimately yielding the elastodynamic result 
\eqref{-13}.

Expressing the effective elastodynamic parameters\ \eqref{-13} via the
matrix $\mathbf{\hat{G}}^{\!\overset{\backslash \!\mathbf{0}}{}}$ ensures
that they are generally regular on the Bloch branches. This is an important
feature that deserves further scrutiny. By analogy with (\ref{B}), denote
the manifold of Bloch branches for vector waves by $\mathbf{\Omega }_{%
\mathrm{B}}$ (bold-lettered to distinguish it from $\Omega _{\mathrm{B}}$ of
the scalar case), so that 
\begin{equation}
\mathbf{\Omega }_{\mathrm{B}}\ni \left( \omega ,\mathbf{k}\right) _{\mathrm{B%
}}:\ \det \mathbf{\hat{Z}}=0\ \Leftrightarrow \mathbf{\hat{Z}\hat{u}}_{%
\mathrm{B}}=\mathbf{0},  \label{Bv}
\end{equation}%
where the matrix $\mathbf{\hat{Z}}$ and the vector $\mathbf{\hat{u}}_{%
\mathrm{B}}$ have components $\hat{Z}_{ik}\left[ \mathbf{g},\mathbf{g}%
^{\prime }\right] $ and $\hat{u}_{k}^{\left( \mathrm{B}\right) }\left( 
\mathbf{g}^{\prime }\right) $ in $V_{\mathbf{g}}\times \mathbb{C}^{3}.$ If $%
\det \mathbf{\hat{Z}}^{\!\overset{\backslash \!\mathbf{0}}{}}\neq 0$ and so $%
\mathbf{\hat{G}}^{\!\overset{\backslash \!\mathbf{0}}{}}=\mathbf{\hat{Z}}%
^{\backslash \mathbf{0}^{-1}}$ is well-defined (see\ \eqref{-13}), then 
\begin{equation}
\det {\mathbf{\hat{Z}}}\left( \omega ,\mathbf{k}\right) =\det \langle 
\mathbf{\hat{G}}\rangle ^{-1}\det \mathbf{\hat{Z}}^{\!\overset{\backslash \!%
\mathbf{0}}{}},\ \ \ \langle \mathbf{\hat{G}}\rangle ^{-1}=\langle \mathbf{%
\hat{Z}}\rangle -{\mathbf{\hat{w}}}^{+}\hat{\mathbf{G}}^{\!\overset{%
\backslash \!\mathbf{0}}{}}{\mathbf{\hat{w}\ }}\ \ \mathrm{with\ }\det
\langle \mathbf{\hat{G}}\rangle ^{-1}=0\ \mathrm{on\ }\mathbf{\Omega }_{%
\mathrm{B}}.  \label{84v}
\end{equation}%
This is similar to the scalar version (\ref{84aa})-(\ref{84a}) 
with the important distinction that now $\langle \mathbf{\hat{Z}}\rangle
\equiv \mathbf{\hat{Z}}\left[ \mathbf{0,0}\right] =\mathbf{\hat{e}}^{+}%
\mathbf{\hat{Z}\hat{e}}$ and $\langle \mathbf{\hat{G}}\rangle \equiv \mathbf{%
\hat{G}}\left[ \mathbf{0,0}\right] =\mathbf{\hat{e}}^{+}\mathbf{\hat{G}\hat{e%
}}$ are $3\times 3$ matrices in $\mathbb{\mathbb{C}}^{3}$ and $\mathbf{\hat{w%
}}\in \mathbb{\mathbb{C}}^{3}\otimes V_{\mathbf{g}\neq \mathbf{0}}$, ${%
\mathbf{\hat{e}}}$ $\in \mathbb{\mathbb{C}}^{3}\otimes V_{\mathbf{g}=\mathbf{%
0}}$. Their components in $\mathbb{\mathbb{C}}^{3}$ are defined by 
\begin{equation}
\begin{aligned} 
\langle \mathbf{\hat{Z}} \rangle :\  \langle\hat{Z} \rangle
_{ik} &={k_{j}k_{l}\langle C_{ijkl}\rangle -\omega ^{2}\langle \rho
}_{ik}{\rangle -\omega k_{j}\langle S_{ijk}+S_{kji}^{\ast }\rangle ,} 
 \\ 
\langle \mathbf{\hat{G}} \rangle ^{-1} \equiv \mathbf{Z}^{\text{eff}} :\ 
{Z}_{ik}^{\text{eff} } &=\langle\hat{Z} \rangle _{ik}
-\sum\nolimits_{\mathbf{g},\mathbf{g}^{\prime }\in \Gamma \backslash
\mathbf{0}}\hat{w}{_{ip}^{\ast }}\left( \mathbf{g}\right)
\hat{G}_{pq}^{\!\stackrel{\backslash\!\mathbf{0}}{}}\left(
\mathbf{g,g}^{\prime }\right) \hat{w}_{q k}\left( \mathbf{g}^{\prime
}\right) , 
\\ \mathbf{\hat{w}},\,\mathbf{\hat{e}} :\ \hat{w}_{ik} &= k_j
\hat{q}_{ijk} - \omega \hat{r}_{ik} ,\ \  \hat{e}_{ik} = \delta_{ik}
\delta_{\mathbf{g} \mathbf{0} } \end{aligned}  \label{85}
\end{equation}%
(the quantity $\mathbf{Z}^{\text{eff}}$ is expressed in more physical
terms in \S \ref{imp}). 
Note also that the vector-case generalization of \eqref{84b} implies that the null vector of 
$\langle \mathbf{\hat{G}} \rangle ^{-1} $ $(= \mathbf{Z}^{\text{eff}})$ 
on $\mathbf{\Omega }_{\mathrm{B}}$ is equal to the average $\langle\mathbf{u}_{\mathrm{B}}\rangle$ of the Bloch eigenmode defined by \eqref{Bv}$_2$.
Finally, consider the manifold $\mathbf{\Omega }_{%
\mathrm{exc}}$ of exceptional points where $\det \mathbf{\hat{Z}}^{\!\overset%
{\backslash \!\mathbf{0}}{}}=0$ (cf.\ (\ref{ex1})) and the explicit
expressions \eqref{-13} diverge. For the points in $\mathbf{\Omega }_{%
\mathrm{exc}}\not{\cap}\,\mathbf{\Omega }_{\mathrm{B}}$, i.e.\ lying off the
Bloch branches, we have $\det \langle \mathbf{\hat{G}}\rangle =0$ 
(cf.\ \eqref{84v}$_1$) 
and hence the effective force $\mathbf{f}^{\mathrm{eff}}(\mathbf{x,}%
t)$ (\ref{3-2}) with $\left\langle \mathbf{f}\right\rangle \in \mathbb{%
\mathbb{C}}^{3}$ being a null vector of $\langle \mathbf{\hat{G}}\rangle $
produces a zero effective response $\left\langle \mathbf{u}\right\rangle =%
\mathbf{0.}$ This is again similar to the scalar case (except that the
vector $\left\langle \mathbf{f}\right\rangle $ has to be specialized while $%
\left\langle u\right\rangle =0$ $\forall \,\left\langle f\right\rangle $).
If the intersection $\mathbf{\Omega }_{\mathrm{exc}}\cap \mathbf{\Omega }_{%
\mathrm{B}}$ occurs, i.e.\ $\det \mathbf{\hat{Z}}^{\backslash \mathbf{0}}=0$
occurs on a Bloch branch, then, contrary to the scalar case, the vector
Bloch mode does not generally have a zero mean 
$\langle \mathbf{u}_{\mathrm{B}}\rangle =\mathbf{0}$ unless the null vector $\mathbf{\hat{h}%
}$ of $\mathbf{\hat{Z}}^{\!\overset{\backslash \!\mathbf{0}}{}}$ also
satisfies $\mathbf{\hat{w}}^{+}\mathbf{\hat{h}=0}$ (see eq.\ \eqref{83}$_{1}$%
).

\section{Discussion of the PWE effective model}

\label{sec5}

\subsection{Properties of the effective constants}

\label{4.1}

As it was noted in \S \ref{sum1}, the effective material tensors given by
eqs.\ (\ref{-13}) retain the symmetries (\ref{-434}) of the corresponding
tensors of the initial periodic material. At the same time, the effective
density and elasticity tensors ${\pmb\rho }^{\text{eff}}$ (at $\omega
\neq 0$) and $\mathbf{C}^{\text{eff}}$ are not sign definite.  
This is not at all surprising since (\ref{=65}) no longer describes a positive energy for the homogenized dispersive medium. 
 Dispersive effective tensors 
defined by (\ref{-13}) correspond to the non-local properties of the
homogenized medium in the space-time domain where $\mathbf{C}^{\text{eff}},~%
{\pmb\rho }^{\text{eff}}$ and $\mathbf{S}^{\text{eff}}$ depend on $%
\mathbf{\xi }=\mathbf{x-x}^{\prime }$ and $\tau =t-t^{\prime }$. Note that
the effective tensors are certainly not periodic in $\mathbf{k}$ (even if
taken on the periodic Bloch branches $\omega _{\mathrm{B}}\left( \mathbf{k}%
\right) =\omega _{\mathrm{B}}\left( \mathbf{k+g}\right) $), and hence they
are not periodic in $\mathbf{\zeta ,}$ as well as non-stationary.
It is therefore clear that the initial periodic medium with non-zero
coupling $\mathbf{S}\left( \mathbf{x}\right) $ in (\ref{=64}) cannot be seen
as resulting from PWE homogenization of another periodic material, say with
a much finer scale. At the same time, by proceeding from (\ref{=64}) and
arriving at (\ref{00}) with the same entries of the coupling terms $\mathbf{S%
}$ and $\mathbf{-S}^{+}$, the PWE homogenization confirms the closure of the
Willis elastodynamics model. In this regard, note that  our notation $\mathbf{S}%
^{+}$ defined as $S_{\left( ij\right) k}^{+}\equiv S_{k\left( ij\right)
}^{\ast }$ is equivalent to $\mathbf{S}^{\dagger }$ of \cite{Nasser11b} (see
also the footnote below).

Consider in more detail the conventional situation where the initial
periodic medium represents a classically elastic solid with zero coupling $%
\mathbf{S}\left( \mathbf{x}\right) =0$ and real-valued density $\rho \left( 
\mathbf{x}\right) $ and stiffness $\mathbf{C}\left( \mathbf{x}\right) $. It
is evident from (\ref{-13}) with $\mathbf{\hat{S}}=0$ that ${\pmb\rho }^{%
\text{eff}}$ and $\mathbf{C}^{\text{eff}}$ are even functions and $\mathbf{S}%
^{\text{eff}}$ an odd function of $\omega $ and hence of $\tau .$ Dependence on 
$\mathbf{k}$ satisfies $\omega _{\mathrm{B}}\left( \mathbf{k}\right) =\omega
_{\mathrm{B}}\left( -\mathbf{k}\right) $ and 
\begin{equation}
{\pmb\rho }^{\text{eff}}\left( \omega ,\mathbf{k}\right) ={\pmb\rho }^{%
\text{eff}^{\ast }}\left( \omega ,-\mathbf{k}\right) ,\ \mathbf{C}^{\text{eff%
}}\left( \omega ,\mathbf{k}\right) =\mathbf{C}^{\text{eff}^{\ast }}\left(
\omega ,-\mathbf{k}\right) ,\ \mathbf{S}^{\text{eff}}\left( \omega ,\mathbf{k%
}\right) =-\mathbf{S}^{\text{eff}^{\ast }}\left( \omega ,-\mathbf{k}\right) .
\label{5.1}
\end{equation}%
By (\ref{5.1}), $\mathbf{C}^{\text{eff}},~{\pmb\rho }^{\text{eff}}$ are
real and $\mathbf{S}^{\text{eff}}$ is pure imaginary at $\mathbf{k=0}$ and
any $\omega $. It also follows from (\ref{5.1}) that the $\mathbf{C}^{\text{%
eff}}\left( \mathbf{\xi ,}\tau \right) $ and ${\pmb\rho }^{\text{eff}%
}\left( \mathbf{\xi ,}\tau \right) $ are real and $\mathbf{S}^{\text{eff}%
}\left( \mathbf{\xi ,}\tau \right) $ is pure imaginary 
in the space-time domain\footnote{
This observation  makes consistent the use of $-\mathbf{S}^{+}$ in (\ref{=64}) in place of $\mathbf{S}^{\dagger }=\mathbf{S}^{T}$ in \cite[eqs.\ (7.3),
(7.5)]{Milton07}.   Equation \eqref{5.1}$_3$  implies that the term $-{\mathbf{S}^{\text{eff}}}^{+}$ in the homogenized equations \eqref{00} can be interpreted as ${\mathbf{S}^{\text{eff}}}%
^{+}(\omega ,-\mathbf{k})$ which is the formal adjoint of $\mathbf{S}$
considered as a differential operator, i.e.\ $\mathbf{S}^{\dagger }$, see  \cite{Willis09,Willis11}.
The variety of notation  in the literature for the coupling $\mathbf{S}^{\text{eff}}$  does not make things simpler.  In this regard, notation  in \cite{Nasser11b} agrees with \eqref{00} and \eqref{5.1};
notation in \cite[eq.\ (3.26)]{Willis97} is similar to \eqref{00} although
the conjugate of $M_{ijk}$ (equivalent to $S_{ijk}$ here) does not appear.
}.

Let ${\rho }\left( \mathbf{x}\right) $ and $\mathbf{C}\left( \mathbf{x}%
\right) $ be even functions of $\mathbf{x}$. This condition on its own
implies that ${\pmb\rho }^{\text{eff}}$ and $\mathbf{C}^{\text{eff}}$ are
even functions and $\mathbf{S}^{\text{eff}}$ is an odd function of $\mathbf{k}$
and hence of $\mathbf{\xi }$. Combination with (\ref{5.1}) implies that a
classically elastic solid with a periodically even profile is characterized
by ${\pmb\rho }^{\text{eff}}$ and $\mathbf{C}^{\text{eff}}$ that are real
and even (in each variable) functions of either set of variables $\left(
\omega ,\mathbf{k}\right) $ and $\left( \mathbf{\xi ,}\tau \right) ,$ while $%
\mathbf{S}^{\text{eff}}$ is a real valued odd function of $\left( \omega ,\mathbf{k}%
\right) $ which vanishes at $\mathbf{k=0}$ and it is a pure imaginary odd
function of $\left( \mathbf{\xi ,}\tau \right) $.

The case of an even profile is one example of possible symmetry. The general case may
be outlined as follows. Let the orthogonal transformation $\mathbf{R}$ ($
\mathbf{R}^{-1}=\mathbf{R}^{T}$) be an element of the symmetry group  of the initial material, such that  $C_{ijkl}\left( \mathbf{x}\right)
=R_{im}R_{jn}R_{kp}R_{lq}C_{mnpq}\left( \mathbf{x}\right) $  $\forall 
\mathbf{x}$ and also $\rho \left( \mathbf{x}\right) =\rho \left( \mathbf{Rx}%
\right) $ and $\mathbf{C}\left( \mathbf{x}\right) =\mathbf{C}\left( \mathbf{%
Rx}\right)$.  Then $\omega \left( \mathbf{k}\right) =\omega \left( \mathbf{Rk%
}\right) $ and
\begin{equation}\label{5.2'}
\begin{aligned}
 \rho _{ij}^{\text{eff}}\left( \omega ,\mathbf{k}\right) &=R_{im}R_{jn}\rho
_{mn}^{\text{eff}}\left( \omega ,\mathbf{Rk}\right) ,\ 
\\    C_{ijkl}^{\text{eff}%
}\left( \omega ,\mathbf{k}\right) &=R_{im}R_{jn}R_{kp}R_{lq}C_{mnpq}^{\text{%
eff}}\left( \omega ,\mathbf{Rk}\right) ,\  \\ 
S_{ijk}^{\text{eff}}\left( \omega ,\mathbf{k}\right)
&=R_{im}R_{jn}R_{kp}S_{mnp}^{\text{eff}}\left( \omega ,\mathbf{Rk}\right) .%
\end{aligned}
\end{equation}%
Consequently, the effective tensors ${\pmb\rho }^{\text{eff}},$ $%
\mathbf{C}^{\text{eff}}$ and $\mathbf{S}^{\text{eff}}$ are invariant to  
$\mathbf{R}$ at $\mathbf{k=0,}$ and are invariant at $\mathbf{k\neq 0}$
to those orthogonal transformations which also 
 satisfy $\mathbf{Rk=k}$.

\subsection{Effective impedance}     \label{imp}

Assume the homogenized medium with the effective elastodynamics equations in
the Willis form \eqref{00}. The problem of finding the effective response to
effective sources (\ref{3-2}) averaged over the unit cell of the original
periodic medium is now seen as a forced-motion problem:%
\begin{equation}
\{\mathbf{f}^\text{eff},{\pmb\gamma}^\text{eff}\}\left( \mathbf{x,}t\right) 
=\{ \langle\mathbf{f}\rangle,\langle{\pmb \gamma }\rangle\}
\,e^{i(\mathbf{k}\cdot \mathbf{x}-\omega t)}~\Rightarrow ~\mathbf{u}^\text{eff}
\left( \mathbf{x,}t\right) \ = \langle\mathbf{u}\rangle \, e^{i(\mathbf{k}\cdot \mathbf{x}
-\omega t)} .
\label{3-}
\end{equation}%
In particular, uniform amplitudes $\langle\mathbf{f}\rangle$ and $\langle\pmb{\gamma }\rangle$ can now
be interpreted as Fourier harmonics of arbitrary sources in the homogenized
medium. 
Equations \eqref{00} with reference to (\ref{3-}) can be recast as
\begin{equation}
\mathbf{Z}^{\text{eff}}(\omega ,\mathbf{k}) \, \langle\mathbf{u}\rangle 
= \langle\mathbf{f}\rangle +i\big( \omega {
{\mathbf{S}}^{\text{eff}}}^{+} -{\pmb k}^{+}{\mathbf{C}}^{\text{eff}}\big)
\langle\pmb{\gamma }\rangle ,  \label{123}
\end{equation}
where ${\mathbf{Z}}^{\text{eff}}$ is the effective impedance matrix with the
components in $\mathbb{C}^{3}$ 
\begin{equation}
Z_{ik}^{\text{eff}}=C_{ijkl}^{\text{eff}}k_{j}k_{l}-\omega k_{j}(S_{ijk}^{%
\text{eff}}+{S_{kji}^{\text{eff}}}^{\ast })-\omega ^{2}\rho _{ik}^{\text{eff}%
}\ \ \ \left( \Rightarrow \mathbf{Z}^{\text{eff}}=\mathbf{Z}^{\text{eff+}%
}\right) .  \label{124}
\end{equation} 
Equation (\ref{123}) considered in the absence of sources 
implies
\begin{equation}
\mathbf{Z}^{\text{eff}}  \langle\mathbf{u}_{\mathrm{B}}\rangle 
= 0, \quad 
\det \mathbf{Z}^{\text{eff}}(\omega ,\mathbf{k})=0 .  \label{125}
\end{equation}
Equation \eqref{125}$_2$ is the dispersion equation for free waves in the homogenized medium, and as such may be called the Christoffel-Willis equation due to its  analogy to the Christoffel equation in a uniform medium. 
Note that, strictly speaking, ${\mathbf{Z}}^{\text{eff}}$ has been introduced above
as the impedance of the homogenized medium. We now show that it can be
equally interpreted as the effective impedance for the initial periodic
material. The consistency of such dual interpretation of $\mathbf{Z}^{\text{%
eff}}$ is an essential verification of the model.

For clarity, consider first the case of a scalar wave problem, where the
impedance (\ref{124}) is a scalar 
\begin{equation}
Z^{\text{eff}}=\mathbf{k}\cdot {\pmb\mu }^{\text{eff}}\mathbf{k}-\omega 
\mathbf{k}\cdot (\mathbf{S}^{\text{eff}}+{\mathbf{S}^{\text{eff}}}^{\ast
})-\omega ^{2}\rho ^{\text{eff}}  \label{343}
\end{equation}%
and (\ref{125}) reduces to $Z^{\text{eff}}(\omega ,\mathbf{k})=0.$
Substituting the effective material parameters (\ref{03}) in (\ref{343})
yields the identity 
\begin{equation}
{Z}^{\text{eff}}(\omega ,\mathbf{k})=\langle \hat{Z}\rangle -\mathbf{\hat{w}}%
^{+}\mathbf{\hat{G}_{\backslash \mathbf{0}}\hat{w}.}  \label{344}
\end{equation}%
Now recall that $\langle \hat{Z}\rangle -{\mathbf{\hat{w}}^{+}\mathbf{\hat{G}%
_{\backslash \mathbf{0}}\hat{w}}}=\langle \hat{G}\rangle ^{-1}$ by (\ref%
{84aa}), and that, barring certain exceptions (see \S \ref{excep}), $\langle 
\hat{G}\rangle ^{-1}(\omega ,\mathbf{k})=0$ is equivalent to the dispersion
equation (\ref{B}) defining Bloch branches in the actual periodic material.
Thus we arrive at eq.\ (\ref{04}), where ${Z}^{\text{eff}}=\langle \hat{G}%
\rangle ^{-1}$ is now an equality, and we observe that $Z^{\text{eff}%
}(\omega ,\mathbf{k})=0$ is the same dispersion equation providing both the
spectrum of eigenmodes for the homogenized medium and the Bloch spectrum (in
the main) for the actual periodic material. It is instructive to note from (%
\ref{344}) that the effective impedance ${Z}^{\text{eff}}$ generally differs
from 
$\langle \hat{Z}\rangle \equiv {\mathbf{\hat{e}^+\hat{Z}\hat{e}}}$ defined by
the statically averaged material constants (see (\ref{82})). This
observation highlights the fact that the PWE dynamic homogenization of a
periodic material must necessarily proceed from a forced-wave problem.

The same conclusion follows for the elastodynamic case. Substituting from (%
\ref{3-}) into (\ref{124}) yields the same expression as obtained for $%
\langle \mathbf{\hat{G}}\rangle ^{-1}$ in (\ref{85}) and therefore verifies
that $\mathbf{Z}^{\text{eff}}=\langle \mathbf{\hat{G}}\rangle ^{-1}$, now an
equality. By (\ref{84v}), $\det \langle \mathbf{\hat{G}}\rangle ^{-1}=0$ is,
apart from exceptional cases, equivalent to the Bloch dispersion equation.
Comparison with (\ref{125}) corroborates the desired consistency. Note $%
\mathbf{Z}^{\text{eff}}\neq \langle \mathbf{\hat{Z}}\rangle $ in (\ref{85}),
which implies the similar observation as noted above for the scalar case.

The symmetry of the effective material tensors (see \S \ref{4.1}) may lead in a standard
fashion to quasidiagonal or diagonal structure of $\mathbf{Z}^{\text{eff}}.$
To be specific, consider an  example where $\mathbf{k\neq 0}$ is
parallel to one of the orthogonal translations $\mathbf{a}_{j}$ of a 3D simple
cubic lattice of spherical inclusions of locally isotropic material in a
locally isotropic matrix material. Then, using elementary considerations of symmetry,  the tensors 
${\pmb\rho }^{\text{eff}},$ $\mathbf{C}^{\text{eff}}$ and $\mathbf{S}^{%
\text{eff}}$ must satisfy the tetragonal symmetry ($4mm$) and hence $\mathbf{%
Z}^{\text{eff}}$ is diagonal with two equal components$.$ This indicates the
polarization of the averaged Bloch eigenmodes defined by \eqref{125}$_1$ and also
shows that the Bloch spectrum $\omega _{\mathrm{B}}\left( \mathbf{k}\right) $
defined by eq.\ \eqref{125}$_2$ necessarily contains a set of branches that are doubly
degenerate for any non-zero $\mathbf{k\parallel a}_{i}$.

 {A remark is in order concerning the effective 
properties 
for $\mathbf{k=0}$.}
Normally eqs.\ (49$_{2}$) and (50) imply that $\det {\pmb\rho }^{\text{eff}%
}=0$ ($\rho ^{\text{eff}}=0$ in the scalar-wave case) at the 
points $\mathbf{k=0}$, $\omega \left( \mathbf{0}\right) \neq 0$ of the Bloch branches. It is
however noted that $\mathbf{Z}^{\text{eff}}\left( \omega ,\mathbf{0}\right)
=-\omega ^{2}{\pmb\rho }^{\text{eff}}$ may not hold if $\omega \left( 
\mathbf{0}\right) $ belongs to exceptional points of divergence of the
effective tensors (see \S \S 3.4, 4). For example, assume a model case of
periodic $\mathbf{C}\left( \mathbf{x}\right) $ and constant $\rho \equiv
\rho _{0}.$ It is clear that $\mathbf{Z}^{\text{eff}}\left( \omega ,\mathbf{0%
}\right) \neq -\omega ^{2}\rho _{0}\mathbf{I,}$ since the equality would
mean that the Bloch spectrum has no open stopbands at $\mathbf{k=0}$ which
is certainly not the case for an arbitrary periodic $\mathbf{C}\left( 
\mathbf{x}\right)$.  
{In fact, spatially averaging  the pointwise equilibrium equation 
\begin{equation}\label{-3-3}
\rho_0 \omega^2 \mathbf{u} + \text{div}\, {\pmb \sigma} = 0 
\end{equation}
for  $\mathbf{k=0}$ implies that any Bloch mode with $\omega \ne 0$ must  have  zero mean,  
$\langle \mathbf{u}_{\mathrm{B}}\rangle =\mathbf{0}$.  
 In conclusion,} if $\rho $ is constant while $\mathbf{C}\left( 
\mathbf{x}\right) $ is periodic, $\mathbf{k=0}$ definitely belongs to the
exceptional subspace where $\mathbf{C}^{\text{eff}}\left( \omega ,\mathbf{0}%
\right) $ and/or $\mathbf{S}^{\text{eff}}\left( \omega ,\mathbf{0}\right) $
must diverge.  A numerical illustration of this is given in \S\ref{sec6}. 

\subsection{Low frequency limit of the density}  \label{low}

The form of the homogenized effective moduli of \eqref{-13} while relatively
compact belies their intricate dependence on the fixed material properties
of the periodic system as well as on frequency and wave-vector. Here we
touch on one small corner of the parameter space, the quasistatic limit,
with particular attention to the effective inertia. Assume a classical
initial medium (with ${\mathbf{S}}=0$ and positive $\rho (\mathbf{x})$).
Taking $\omega \rightarrow 0,$ it is clear from {\eqref{-13} that to leading
order in $\omega $ the effective moduli 
reduce to their  static form,  
the effective
density is the average, and ${\mathbf{S}}^{\text{eff}} =$O$(\omega )$. The
first correction to the effective
density is positive-definite tensorial, }
\begin{equation}\label{319}
\begin{aligned}
{\pmb\rho }^{\text{eff}}  & =\langle \rho \rangle \,\mathbf{I}+\omega
^{2}\sum\nolimits_{\mathbf{g},\mathbf{g}^{\prime }\in \Gamma \backslash
\! \mathbf{0}}\hat{\rho}(\mathbf{g})\hat{\rho}(-\mathbf{g}^{\prime })\,\hat{\mathbf{G%
}}^{(0)}[\mathbf{g},\mathbf{g}^{\prime }]+\text{o}(\omega ^{2})
\\
& \text{where }\   
\sum\nolimits_{\mathbf{g}^{\prime }\in \Gamma \backslash \! \mathbf{0}}\hat{{G}}%
_{ip}^{(0)}[\mathbf{g},\mathbf{g}^{\prime }]\,\hat{{C}}_{pjkl}(%
\mathbf{g}^{\prime }-\mathbf{g}^{\prime \prime })g_{j}^{\prime
}g_{l}^{\prime \prime }=\delta _{ik}\delta _{\mathbf{g}\mathbf{g}^{\prime
\prime }}.  
\end{aligned} 
\end{equation}%
This illustrates explicitly the departure of the homogenized density from
that of classical elastodynamics, as expected on the basis of prior results
for Willis equations parameters for periodic systems \cite{Shuvalov11,Nasser11b}.

\section{Numerical illustrations and discussion of the homogenization scheme}    
\label{sec6}

\begin{figure}[th]
\begin{center}
\includegraphics[width=3.2in , height=3.0in ]{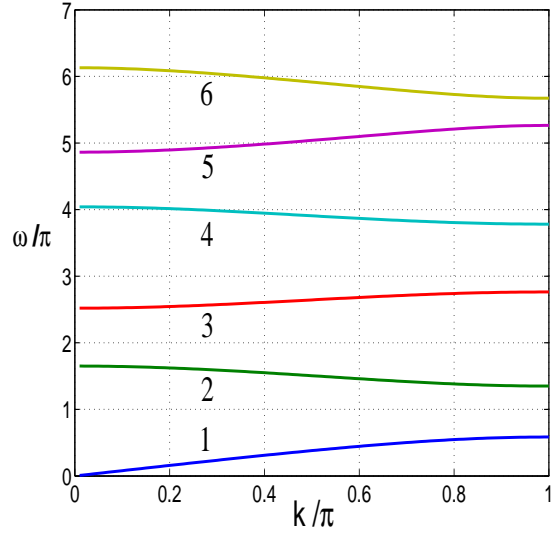}
\end{center}
\caption{The first six Floquet branches for the 1D periodic 
medium with parameters in Table 1.}
\label{fig0}
\end{figure}
\begin{figure}[th]
\centering
\par
\subfigure[First  branch ]{
\includegraphics[width=2.83in , height=2.0in ] {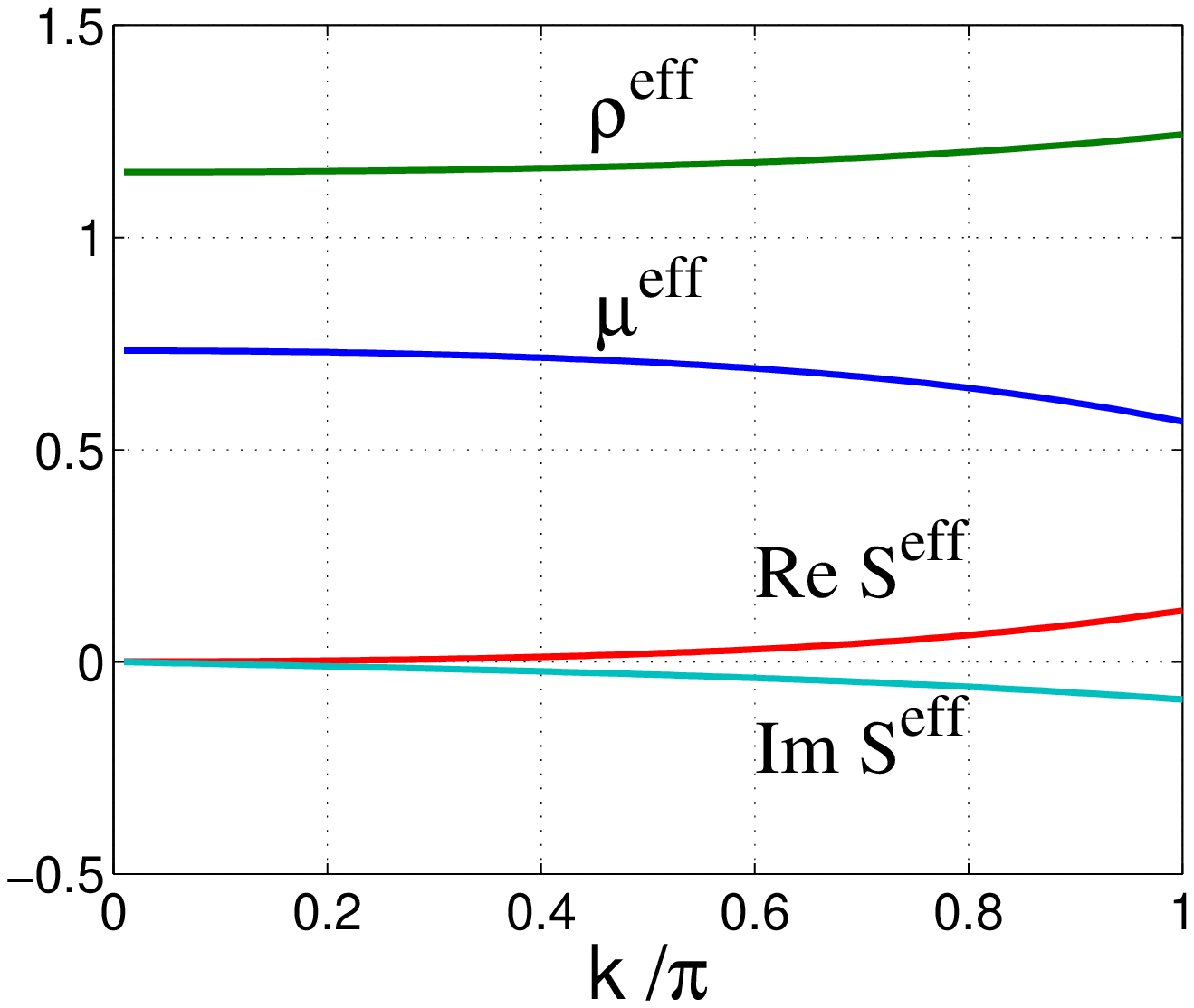}
\label{fig2a}
} 
\subfigure[Second   branch ]{
\includegraphics[width=2.83in , height=2.0in ] {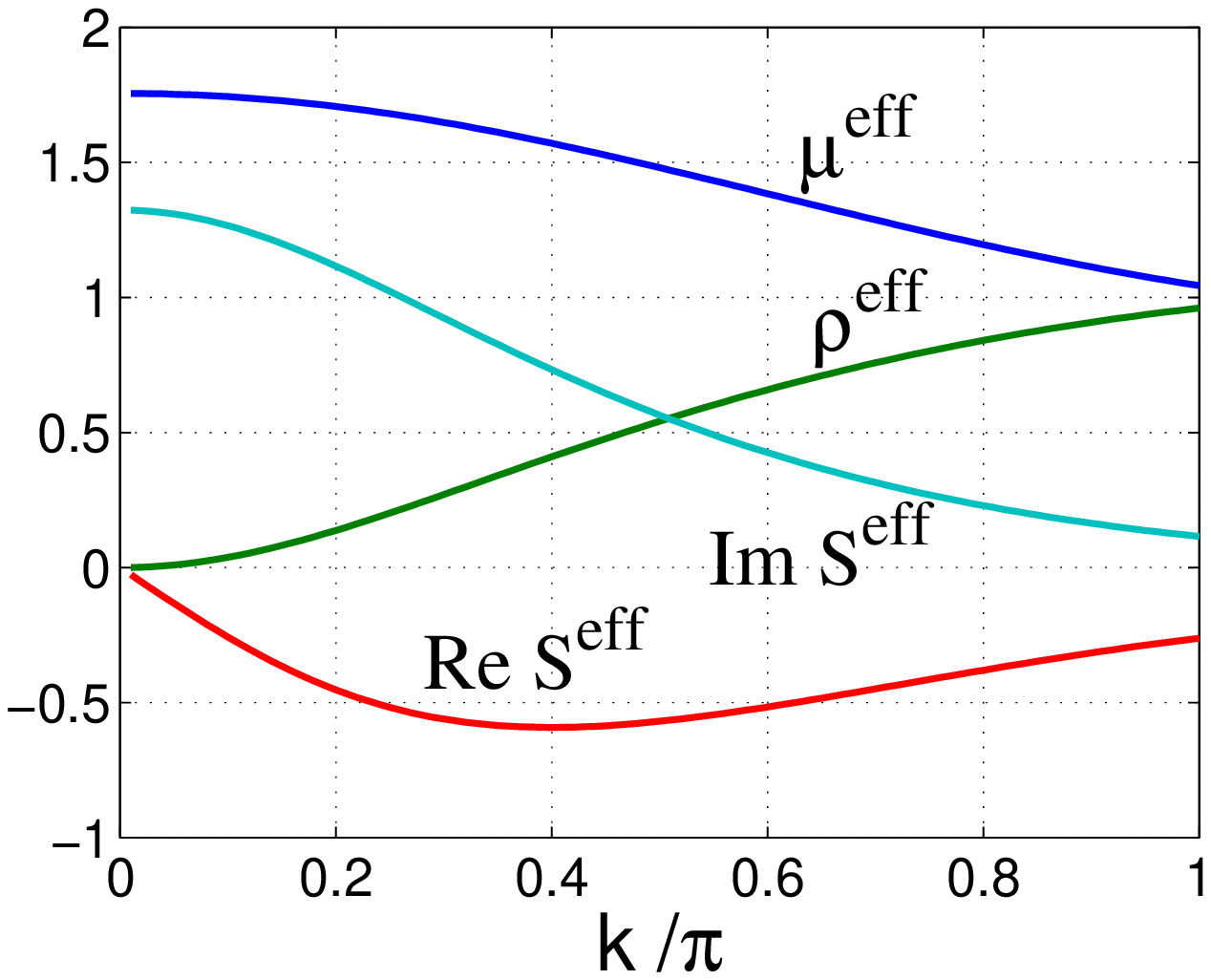}
\label{fig2b}
} 
\subfigure[Third    branch ]{
\includegraphics[width=2.83in , height=2.0in ] {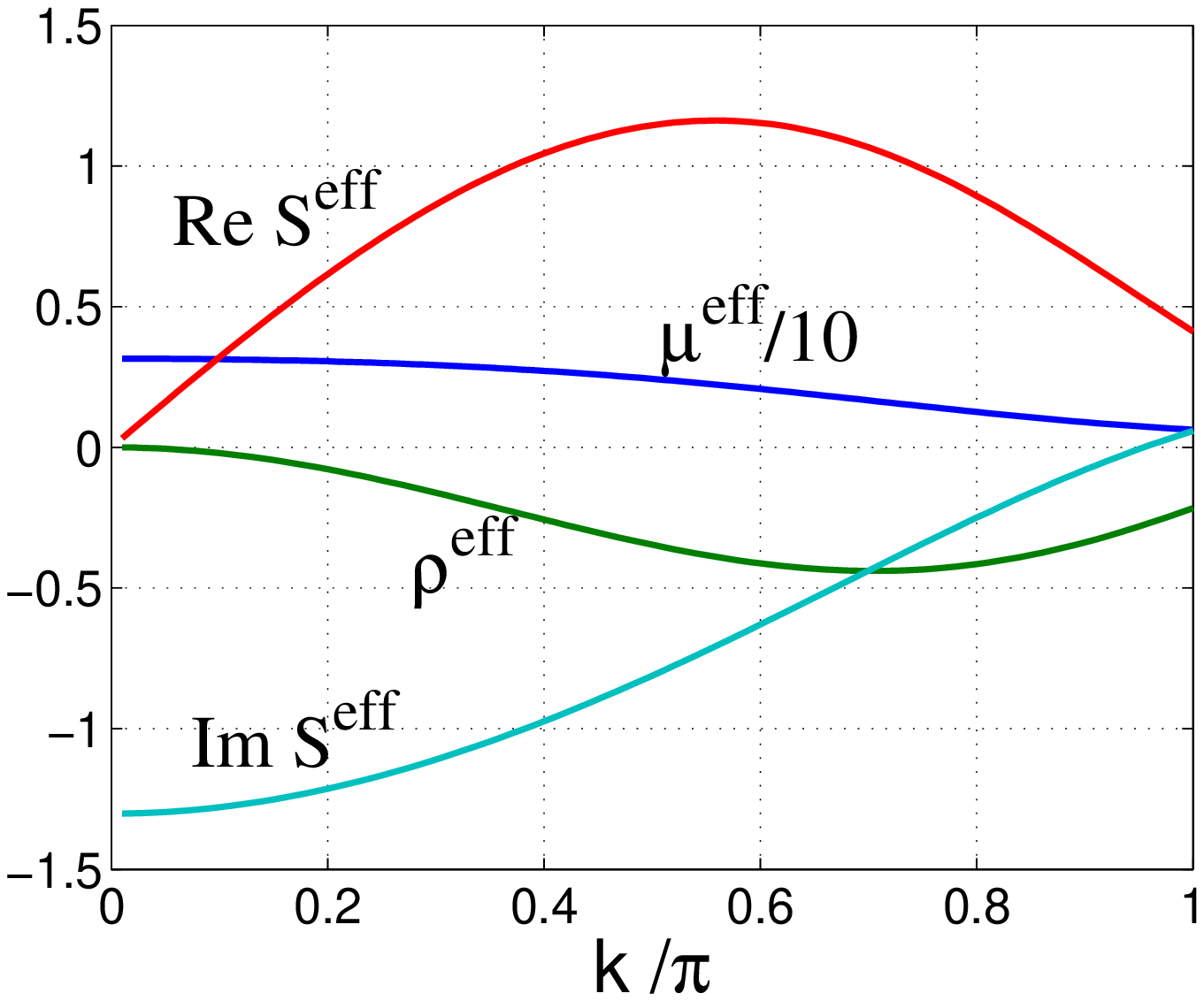}
\label{fig2c}
} 
\subfigure[Fourth   branch ]{
\includegraphics[width=2.83in , height=2.0in ] {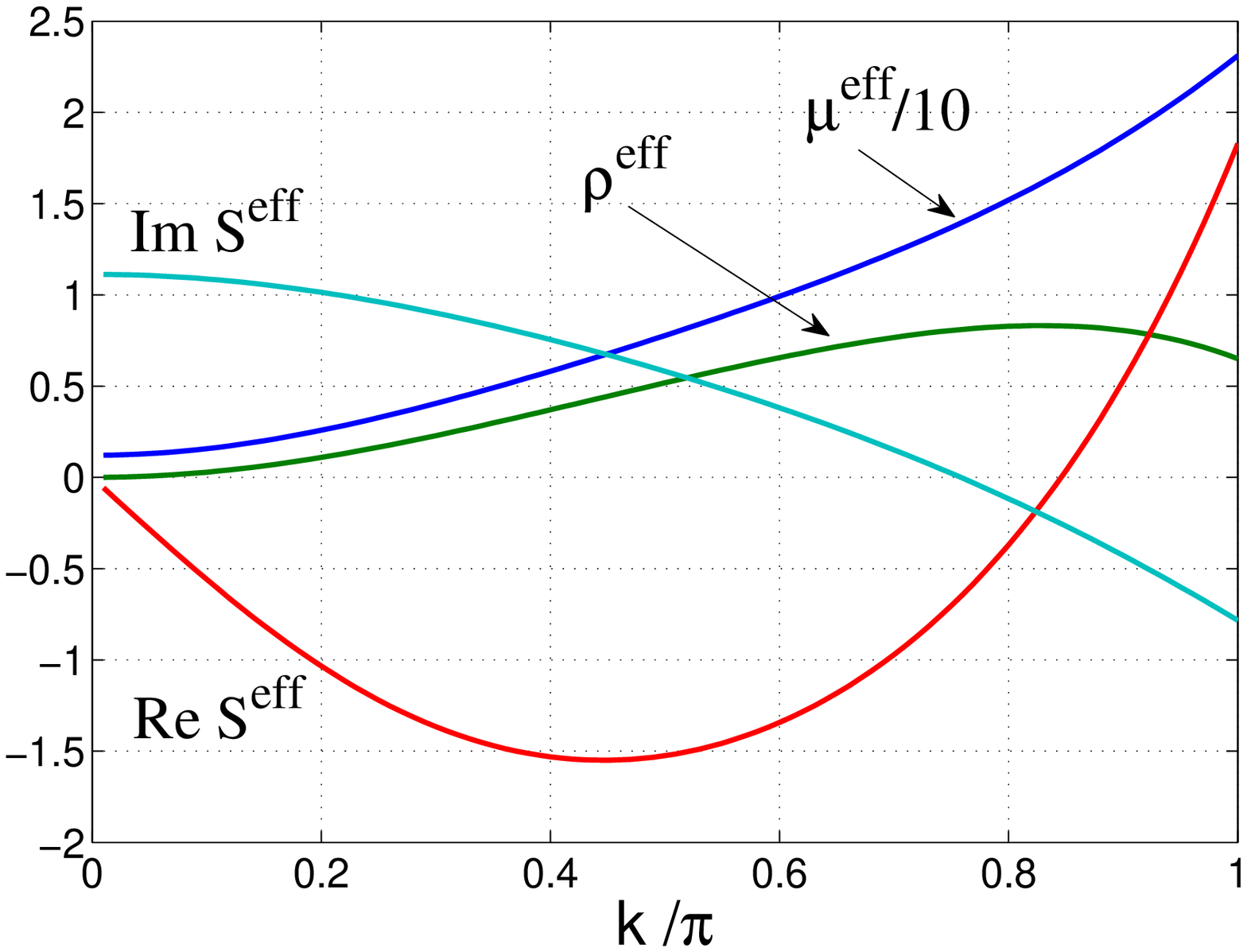}
\label{fig2d}
} 
\subfigure[Fifth    branch ]{
\includegraphics[width=2.83in , height=2.0in ] {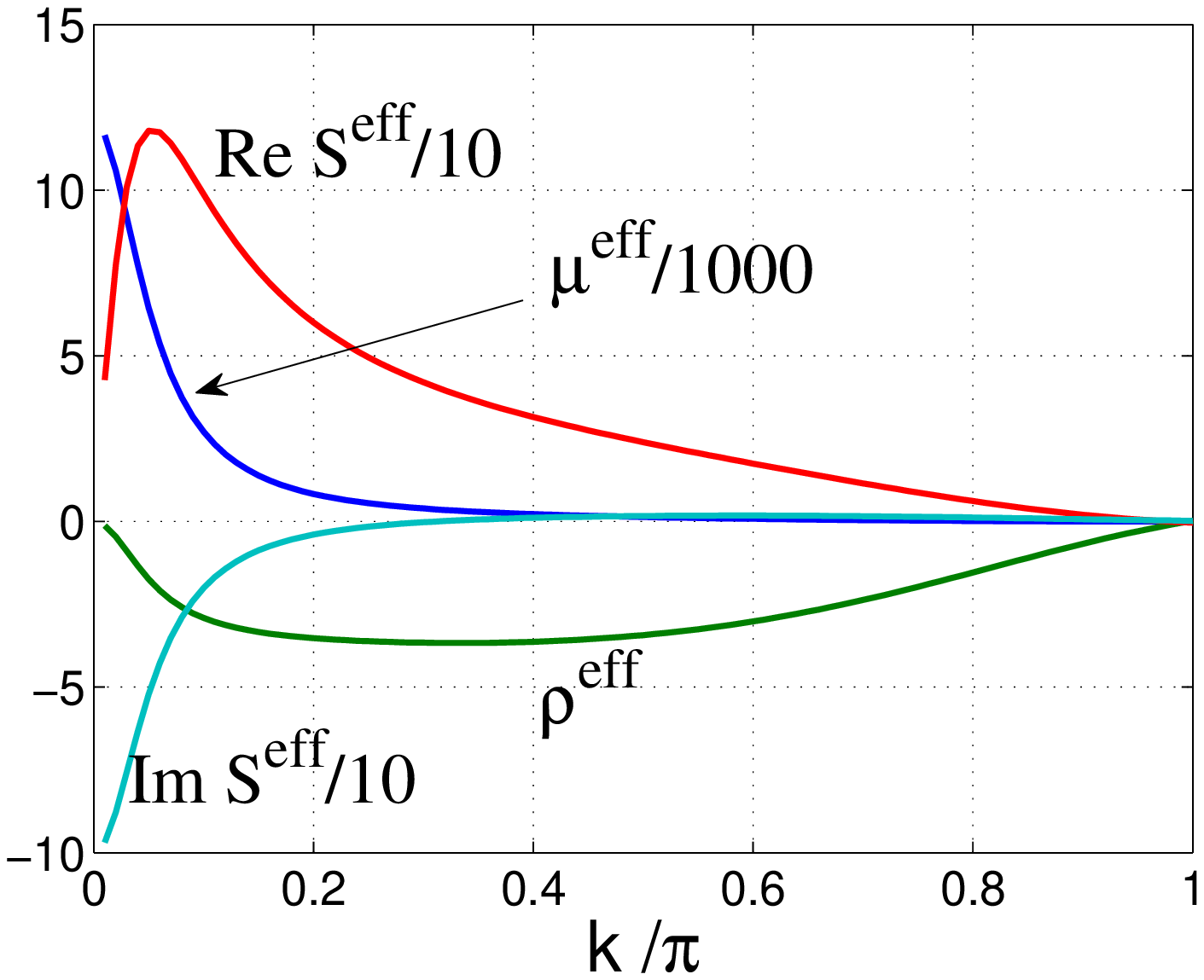}
\label{fig2e}
} 
\subfigure[Sixth   branch ]{
\includegraphics[width=2.83in , height=2.0in ] {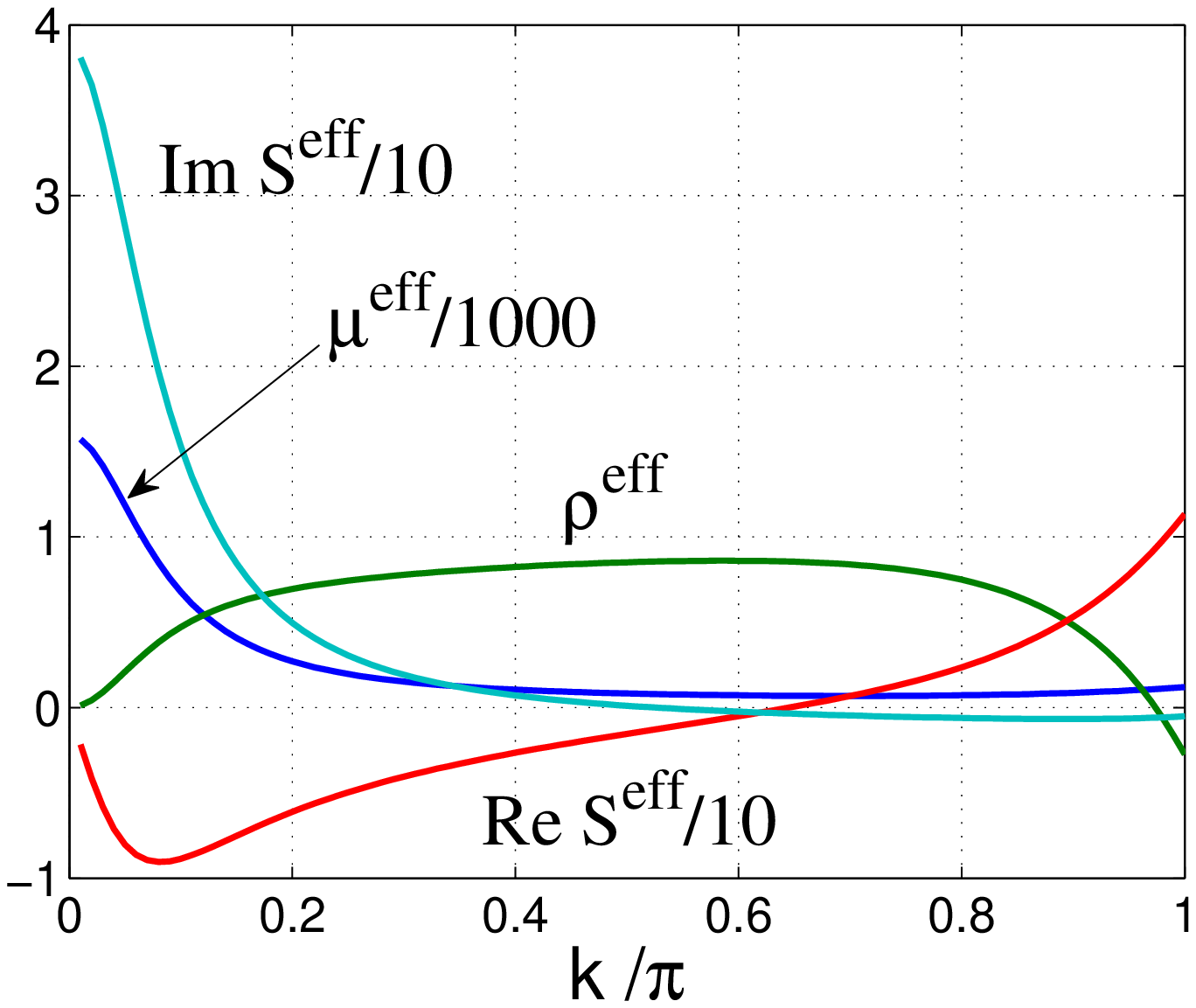}
\label{fig2f}
} \caption{Effective material parameters 
calculated for the 1D periodic medium (Table 1) by the PWE method,  eq.\ 
\eqref{03}.  
The scale is altered on some curves for the sake of legibility. }
\label{fig2}
\end{figure}

Numerical examples are presented for the 1D scalar wave problem. 
All results reported are for the periodic
medium based on the the three-layer single cell defined by Table 1. 
The PWE effective parameters $\mu^\text{eff}$, $\rho^\text{eff}$ and $S^\text{eff}$ are taken on the Bloch branches $\omega_B(k)$ which reduces them to functions of a single variable, $k$. The first six branches $\omega_B(k)$ are shown in Fig.\ \ref{fig0}, and the corresponding values of effective parameters are displayed in Fig.\ \ref{fig2}.

\begin{center}
Table 1: Properties of the three-layer unit cell.  Values were chosen to ensure an unsymmetric cell with non-commensurate layer thicknesses $h$.\\[0pt]
\medskip 
\begin{tabular}{c|ccc}
\hline
layer & $\mu$ & $\rho$ & $h$ \\ \hline\hline
1 & 1 & 1 & 0.37 \\ 
2 & 7 & 2 & 0.313 \\ 
3 & 1/3 & 0.5 & 0.317 \\ \hline\hline
\end{tabular}%
\\[0pt]
~ \\[0pt]
\end{center}
\afterpage{\clearpage}

Setting apart the obvious case of the origin of the first branch which starts
with $S^{\text{eff}}=0$ and $\rho ^{\text{eff}}=\left\langle \rho
\right\rangle $ at $k=0$ (see \S \ref{low}), several other features are noteworthy
in Fig.\ \ref{fig2}. In accordance with (\ref{5.1}), $\text{Re}S^{\text{eff}}$
vanishes at $k=0$ on all branches, while $\text{Im}S^{\text{eff}}$ starts at 
$k=0$ with non-zero values that alternate in sign beginning with the
positive value on the second branch. The effective density $\rho ^{\text{eff}}$ vanishes at $k=0$ on all branches above the first one, as expected from the form of     
the dispersion relation   $Z^\text{eff}(\omega , k)$ = 0, which becomes,  using \eqref{343}, 
\begin{equation}  \label{453}
k^2 \mu^\text{eff} -\omega k(S^\text{eff}+{S^\text{eff}}^*) -\omega^2 \rho^%
\text{eff} =0.
\end{equation}
As $k$ grows, $\rho ^{\text{eff}}\left( k\right) $ exhibits both positive and  negative values, following the alternating trend shown by  Im$S^\text{eff}$.  The magnitude of $\rho^\text{eff}$ remains within the range of the constituent material densities, with the exception of the fifth branch on which it is negative with $|\rho^\text{eff}| > \rho_\text{max}$ for a range of $k$ where $\rho_\text{max}$ is the largest density in the periodic medium ($\rho_\text{max}=2$).  The  calculated PWE  $\mu^\text{eff}$ is always non-negative, however,  its  magnitude  is exceedingly large near $k=0$ for higher order modes, e.g.\ the fifth and sixth. This can be partly understood from the dispersion relation \eqref{453} which  permits $\mu^\text{eff}$ to become large in magnitude
as $k\rightarrow 0$. The  enormous  values on some branches, e.g.\ $\mu^\text{%
eff} > 10,000$ in Fig.\ \ref{fig2e}, may indicate  proximity to an  exceptional point related to a small value of 
$\det \mathbf{\hat{Z}}_{\backslash \mathbf{0}}$ (see \S \ref{excep}).   The magnitude of $S^\text{eff}$ is intermediate between that of the density and the stiffness.    
It is worth noting that  \eqref{=65} evaluated for the 1D scalar Willis model
$( E \rightarrow  \frac 12 ( \mu^\text{eff} k^2 + \rho^\text{eff} \omega^2 )\, |u|^2 )$
 displays negative values on the third and fifth  branches.  The proper form of the energy density for the dispersive effective medium is a delicate topic \cite{Ruppin02}, best left for a separate discussion. 

\begin{figure}[th]
\begin{center}
\includegraphics[width=3.5in , height=3.0in ]{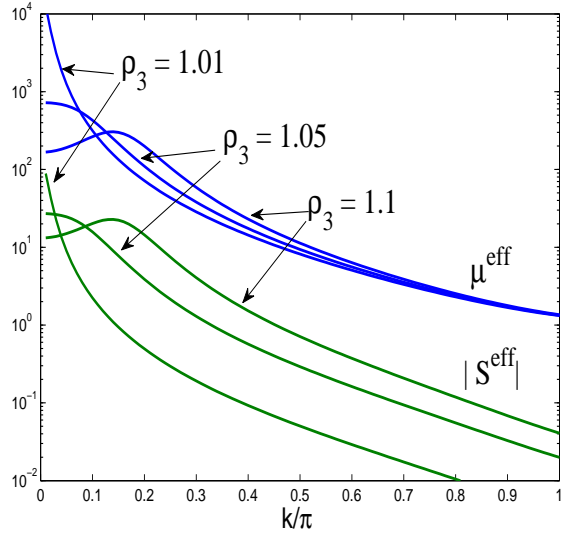}
\end{center}
\caption{The second  Floquet branch for the 1D periodic medium with parameters as  in Table 1,  
except $\rho_2=\rho_1 =1$,  and $\rho_3= (1+\delta) \rho_1$ where $\delta = 1.01$, $1.05$, $1.1$.  The plot shows the developing singular nature  of $\mu^\text{eff}$ and $S^\text{eff}$ at $k=0$ for constant density. }
\label{fig3}
\end{figure}

{Figure \ref{fig3} illustrates the emergence  of  singular effective parameters at $k=0$, $\omega(0) \ne 0$ when the density is uniform, as predicted in \S\ref{imp}.  Although the curves shown are only for the second branch,  the same dependence  is observed in all higher branches.  
}  


An alternative route to homogenization of 1D-periodically stratified elastic
body was proposed in \cite{Shuvalov11}. The basic idea is to compare the logarithm of the
monodromy matrix(MM) $\mathbf{M}\left( y+T,y\right) $, which is the
propagator over one period $T$ along the stratification direction $Y$, with
the matrix of coefficients of the system of elastodynamics equations for a
uniform medium. In this way one can identify the dynamic effective material
parameters at any frequency wave-vector on a Bloch branch as long as the
logarithm is well defined. Unlike the PWE, the MM method does not require
introducing force terms and is not based on averaging. It is worth
emphasizing that the PWE and MM approaches are not two 'technically'
dissimilar derivations of the same result but imply different models of a
homogenized medium. In this light, the more significant is the result that
the MM method also necessarily leads to the effective elasticity of Willis
form (classical elasticity will not suffice) \cite{Shuvalov11}. Note that the MM
homogenization involves the choice of the reference point $y=y_{0}$ at which
the initial data for the monodromy matrix $\mathbf{M}\left(
y_{0}+T,y_{0}\right) $ is prescribed. The MM homogenized media associated
with different $y_{0}$ have the same eigenfrequency spectrum $\omega \left(
k\right) $ which by definition coincides with the Bloch branches of the
actual periodic medium. At the same time, the values of MM effective
constants depend on the choice of $y_{0}$. Once any $y_{0}$ is fixed, the
exact displacement and traction are recovered at periodically distant points 
$y_{0}+nT$. As a result, the MM homogenization can provide an exact solution
of boundary-value or reflection problems for a finite structure containing
several periods or for a periodic half-space in contact with other regions,
see \cite{Shuvalov11}.

\begin{figure}[th]
\centering
\par
\subfigure[$y_0=0 $, at edge of first layer ]{
\includegraphics[width=2.83in , height=2.5in ] {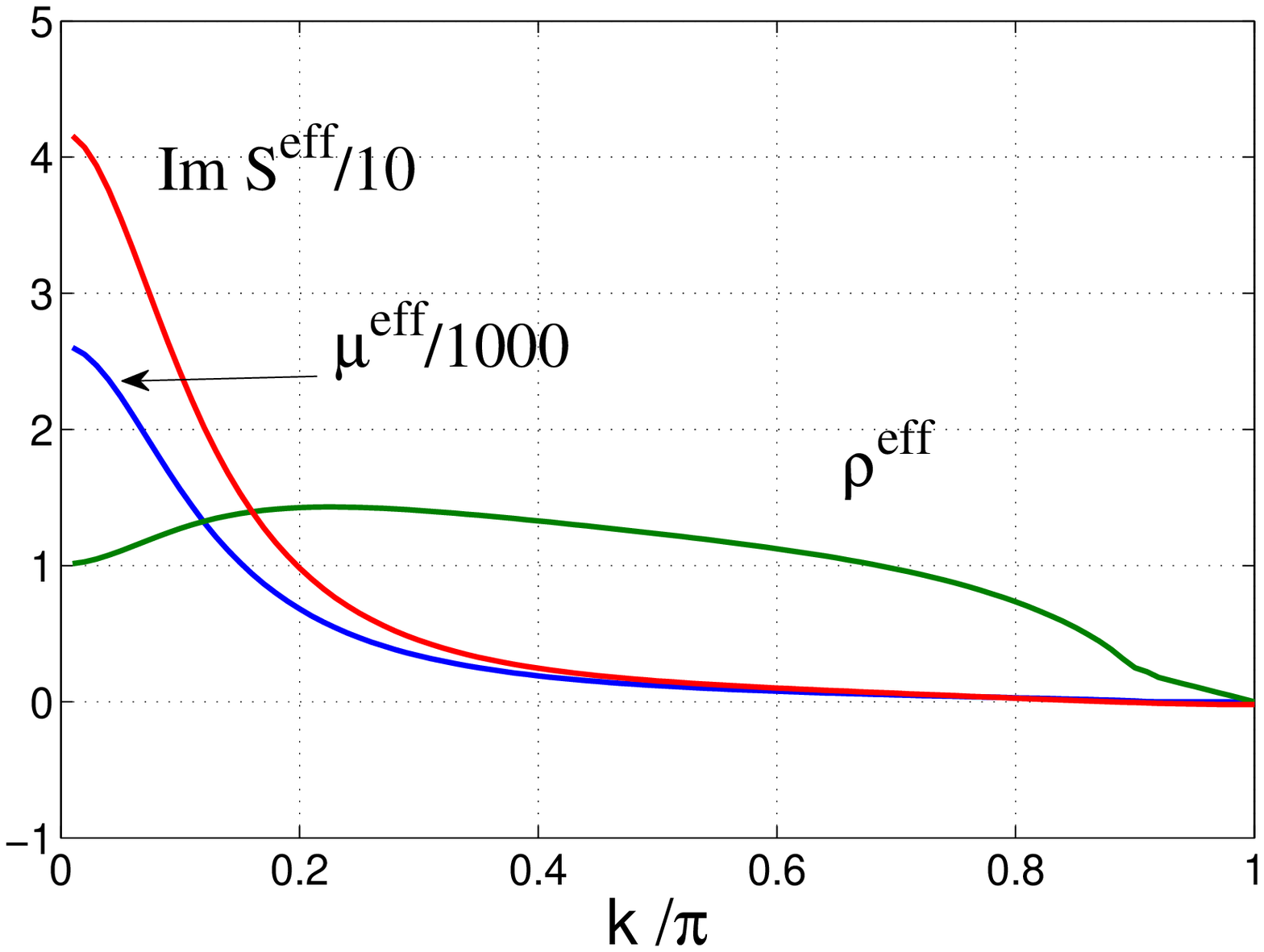}
\label{fig4a}
} 
\subfigure[$y_0=h_1$, at second layer ]{
\includegraphics[width=2.83in , height=2.5in ] {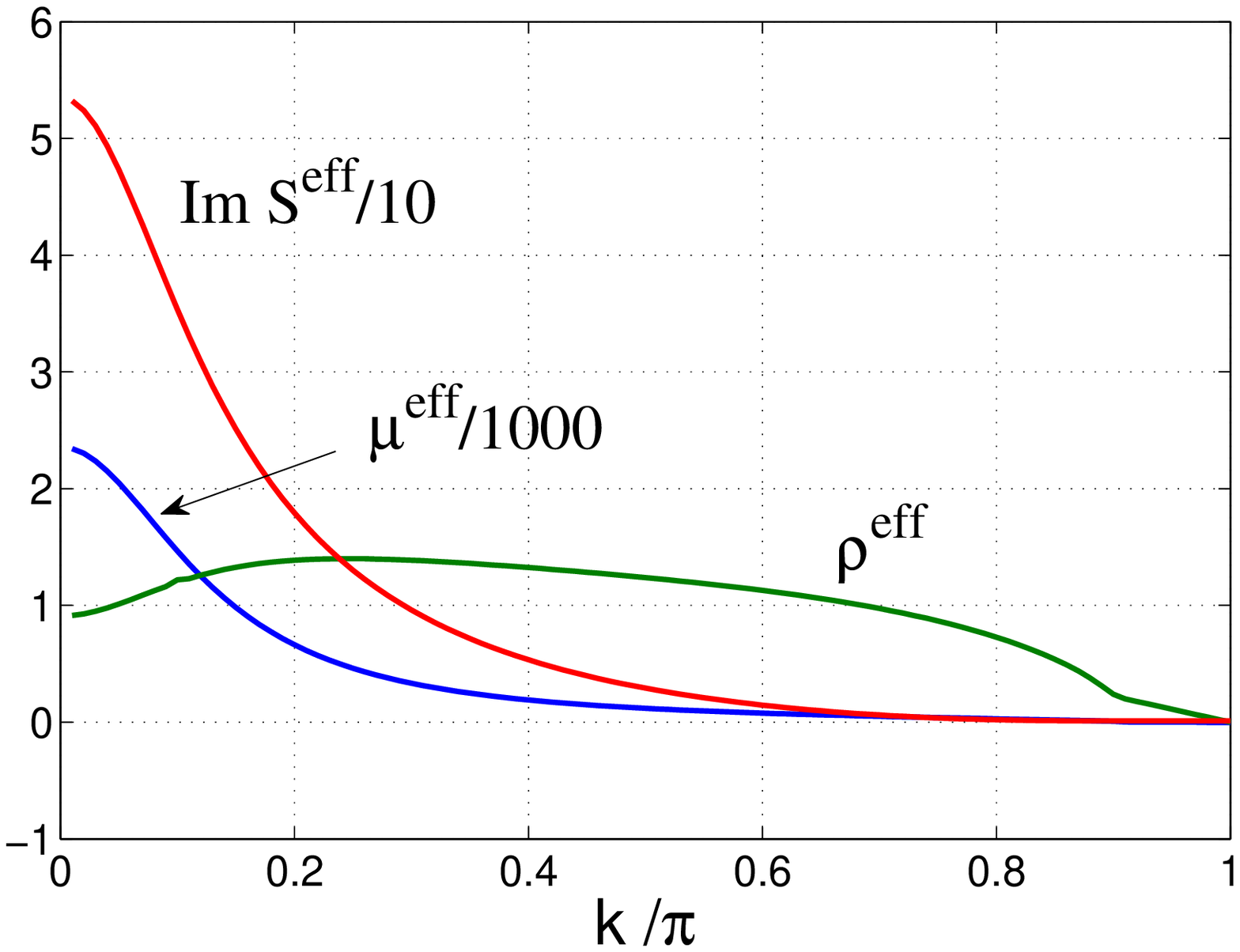}
\label{fig4b}
} 
\caption{Effective material parameters of the 1D periodic medium (Table 1) calculated by the MM method for the fifth Bloch branch. The monodromy matrix ${\mathbf M}(y_0+T,y_0)$ in (a) and (b) is referred to two different locations of the initial point $y_0$ within the unit cell. 
}
\label{fig4}
\end{figure}

As an example, Fig. \ref{fig4} shows the effective parameters of the same material
(Table 1) which are calculated for one of the Bloch branches (the fifth) by
the MM method. The calculation uses the formulas (4.2) and (4.7)$_{1-3}$ of
\cite{Shuvalov11} with a standard definition of the 2$\times $2 propagator $\mathbf{M}\left(
y_{0}+T,y_{0}\right) $ of scalar waves through a trilayer. The two sets of data on Figs.
\ref{fig4a} and \ref{fig4b} correspond to two different choices of the reference point $y_{0}$
within a period. Both sets reveal the same general features of MM
homogenization \cite{Shuvalov10c,Shuvalov11} such as vanishing of $\rho ^{\text{eff}}$ and $\mu^{\text{eff}}$ at $k=\pi $ and a pure imaginary value of $S^{\text{eff}}$. 
The latter stands in stark contrast to the generally complex-valued $S^{\text{eff}}$ in the PWE homogenization model. At the same time, despite the
quite disparate natures of the PWE and MM methods, it is noteworthy that
they do reproduce the enormous values of $\mu^{\text{eff}}$ occurring at small $k$ on the fifth Bloch branch. This may be
seen as  evidence that such behavior of $\mu^{\text{eff}}$ is not a numerical artifact.

\section{Conclusion} \label{sec7}

The principal result of the paper is a set of explicit formulas for
calculating the homogenized material parameters using PWE expansions of the
underlying elastic moduli and density, eq.\ \eqref{-13}. 
The  homogenized governing equations 
retain important physical properties of the original periodic system. Thus, they provide 
(i) the averaged response  to averaged source(s), and (ii) the eigen-spectrum 
$\omega ( \mathbf{k})$ for  Bloch waves.     The effective 
parameters of eq.\ \eqref{-13} are convergent for virtually all
frequency and wave-vector combinations, including $\{\omega , \, \mathbf{k}%
\} $ pairs on the Bloch wave branches.  The exceptional cases  yield singular values of the effective parameters and correspond to zero-mean Bloch waves for the scalar wave system.

Some comments are in order on the PWE homogenization method and its outcome. 
A nonzero strain source is essential for the
derivation of the effective parameters, although the strain source  may be put to zero after the homogenization is complete.    Inclusion of both body force and strain source  is consistent with the findings of \cite{Willis11} that homogenization procedures based on  Green's functions can lead to under-determined systems unless a full set of sources are included from the outset. 
The PWE homogenization procedure derives equations governing the average of the
periodic part of the Bloch wave solution. 
 The PWE scheme is not the only approach  for defining an effective medium
for periodic system. Another approach which applies for media with 1D periodicity is the monodromy matrix (MM) homogenization of \cite{Shuvalov11}.
The effective governing equations are again of Willis form with explicit
expressions for the effective parameters that are distinct from the PWE
parameters, as the numerical example in \S \ref{sec6} demonstrates.   The fact that different sets of effective Willis equations can be obtained for the same medium points to a non-uniqueness in the  definition of dynamic homogenization.   The MM scheme was originally proposed as a method for layered structures that faithfully retains the exact phase relationship over multiple periods, although at  only one point within the period \cite{Shuvalov11}.    The PWE scheme provides predictions  for averaged quantities and does not yield  pointwise exact solutions.

We have shown  that the homogenization of a classically elastic
medium ($\mathbf{S}=0$) yields a Willis effective medium with either complex-valued  
$\mathbf{S}^{\text{eff}}(\omega ,\mathbf{k})$ (satisfying (\ref{5.1})) or
with pure imaginary $\widetilde{\mathbf{S}}^{\text{eff}}(\omega )=- 
\widetilde{\mathbf{S}}^{\text{eff}}(-\omega )$ (\S \ref{sec6}). 
The results indicate the closure property for  the Willis model 
 with strictly complex $\mathbf{S}$   under   homogenization.
 
The present formulation, in starting with a heterogeneous set of Willis elastodynamic equations, and resulting after homogenization in a set of homogeneous constitutive equations of the same form, confirms the  completeness of the Willis model.  
This type of  closure  complements the fact that the Willis constitutive
equations are  closed under spatial transformation \cite{Milton06}.  
Finally,  we note that the present results only partially open the door on our understanding of the effective dynamic properties of phononic crystals. Much remains to be investigated regarding analytical aspects, numerical implementation and applications of these properties.



\end{document}